\documentclass[conference]{IEEEtran}
\IEEEoverridecommandlockouts
\usepackage{cite}
\usepackage{amsmath,amssymb,amsfonts}
\usepackage{algorithmic}
\usepackage{graphicx}
\usepackage{textcomp}
\usepackage{subfig}
\usepackage{xcolor}
\def\BibTeX{{\rm B\kern-.05em{\sc i\kern-.025em b}\kern-.08em
    T\kern-.1667em\lower.7ex\hbox{E}\kern-.125emX}}
\begin{document}

\title{Diffie-Hellman in the Air: A Link Layer Approach\\ for In-Band Wireless Pairing\\
}

\author{
\IEEEauthorblockN{Wenlong Shen\IEEEauthorrefmark{1},
	Yu Cheng\IEEEauthorrefmark{1},
	Bo Yin\IEEEauthorrefmark{1},
	Kecheng Liu\IEEEauthorrefmark{1} and
	Xianghui Cao\IEEEauthorrefmark{2}}
\IEEEauthorblockA{\IEEEauthorrefmark{1} Department of Electrical and Computer Engineering,
	Illinois Institute of Technology, USA}
\IEEEauthorblockA{\IEEEauthorrefmark{2} School of Automation, Southeast University, China}

%\and
%\and
%\and
%\and
}

\maketitle

\begin{abstract}
Key establishment is one fundamental issue in wireless security. The widely used Diffie-Hellman key exchange is vulnerable to the man-in-the-middle attack. This paper presents a novel in-band solution for defending the man-in-the-middle attack during the key establishment process for wireless devices. Our solution is based on the insight that an attacker inevitably affects the link layer behavior of the wireless channel, and this behavior change introduced by the attacker can be detected by the legitimate users. Specifically, we propose a key exchange protocol and its corresponding channel access mechanism for the protocol message transmission, in which the Diffie-Hellman parameter is transmitted multiple times in a row without being interrupted by other data transmission on the same wireless channel. The proposed key exchange protocol forces the MITM attacker to cause multiple packet collisions consecutively at the receiver side, which can then be monitored by the proposed detection algorithm. The performance of the proposed solution is validated through both theoretical analysis and simulation: the proposed solution is secure against the MITM attack and can achieve an arbitrarily low false positive ratio. This proposed link layer solution works completely in-band, and can be easily implemented on off-the-shelf wireless devices without the requirement of any special hardware.
\end{abstract}


\section{Introduction}
With the explosive growth of mobile devices, wearable sensors, smart home appliances, and many other Internet of things, people are living in a world of wirelessly connected devices. Securing the data communication between these wireless devices is of critical importance, especially when sensitive personal data is involved. Cryptographic solutions can be implemented to protect the data communication, however, how to distribute the cryptographic key in the first place is a nontrivial task. It is known that the Diffie-Hellman (DH) key agreement allows two parties with no pre-shared knowledge to jointly establish a shared secret key over the public channel\cite{diffie1976new}. Although the DH protocol is secure against eavesdroppers, its lack of mutual authentication makes it vulnerable to the man-in-the-middle (MITM) attack. The typical approach to address the MITM attack is to execute a device pairing protocol which usually utilizes an out-of-band (OoB) channel, and the utilization of the OoB channel inevitably involves human interactions\cite{mirzadeh2014secure}. Besides the inconvenience brought by the human efforts, the utilization of the OoB channels is also restricted by the requirement of certain user interfaces or device hardware, such as a keyboard or screen. With all these limitations of the OoB channel, it is of significant importance to seek in-band solutions for the initial trust establishment between wireless devices.  
%
%Although a device pairing protocol allow two users to securely establish a common secret rather than negotiating a pre-shared password, its utilization of the OoB channels makes the user experience sometimes no better than carefully selecting and distributing a secure enough password. There have been a few attempts to design in-band solutions for message integrity protection ~\cite{}. However, these solutions require special signal modulation method, which is difficult to be implemented on off-the-shelf wireless devices due to operating system and hardware limitations.  How to design an applicable in-band solution for initial trust establishment for wireless network is still an open issue, and in this paper we will deal with this fundamental challenge.

The necessity of the OoB channel in a device pairing protocol is related to the Dolev-Yao attack model~\cite{dolev1983security}, in which an attacker has full control of the channel and can overhear, intercept, and synthesize any message. However, in a realistic wireless network, an attacker cannot arbitrarily modify an ongoing signal transmission, nor can he cancel the wireless signal propagation~\cite{vcapkun2008integrity}. This realistic adversary model gives people an opportunity to design an in-band solution to deal with the MITM attack during the key exchange process. There have been a few attempts for the in-band device pairing~\cite{gollakota2011secure,hou2013chorus,shen2016secure}. The ideas of these works are based on the I-Code technique~\cite{vcapkun2008integrity}, which can protect the integrity of a message payload by modulating the message signal with Manchester coded ON/OFF keying. There are two major limitations of this type of solutions. First, although the I-Code technique protects the integrity of the protocol messages from being modified by the MITM attacker, it cannot prevent an impersonation attacker. To provide authentication, OoB channels are inevitably involved. For example, \cite{gollakota2011secure} uses the WiFi push button configuration (PBC), which requires the user to physically click the button on the device. However, the proposed link layer solution can deal with the impersonation attack using only in-band channels (discussed in section VII). Second, the I-Code technique requires modification to the physical layer signal modulation method, which is not easy to be implemented on off-the-shelf wireless devices. Our link layer method only requires minimal adjustment on the link layer protocol stack.

The insight of our solution is based the link layer behavior monitorability of a wireless channel, i.e., the attacker's behavior inevitably impacts the wireless channel behavior at the link layer, and this link layer behavior change can be observed by legitimate users. Specifically, to launch the MITM attack, an attacker has to replace the original message with his own. Because an attacker cannot arbitrarily modify an ongoing wireless message, he has to intercept the original message and forge a new one. However, since the attacker cannot cancel the signal propagation of the original message, in order to intercept it, the best he can do is to transmit a jamming signal to collide the original message such that the receiver cannot decode it. This message collision introduced by the attacker is perceptible to the legitimate receiver. 

In this paper, we will exploit this link layer behavior monitorability property of the wireless channel to design a novel in-band solution to deal with the MITM attack. To achieve this, our protocol design has the legitimate user transmit the DH key exchange message multiple times in a row, such that a MITM attacker has to jam all these messages, which will lead to a burst sequence of packet collisions at the receiver's antenna\footnote{In this paper, we use ``packet'' to generally mean a chunk of information delivered over the network. We use ``packet'', ``message'', and ``link-layer frame'' interchangeably, for the convenience of presentation. In our protocol, each message is carried by a single link-layer frame.}. This abnormal link layer behavior can be then detected by the detection algorithm at the receiver's side. The details of our proposed scheme are designed specifically for IEEE 802.11 based wireless networks, which is the most widely used wireless communication standard. However, the methodology of our solution can be used for designing key establish protocols for other distributed coordinated contention based wireless networks, such as 802.15.4 networks. The contribution of this paper can be summarized as follows.
\begin{enumerate}
	\item We systematically study the MITM attack over wireless networks, and model the attacker's behavior on the link layer.
	\item We propose a key establishment protocol based on the DH key exchange, and its corresponding channel access mechanism. The proposed protocol forces a successful MITM attacker to cause consecutive packet collisions at the link layer.
	\item We design an attacker detection algorithm, which can distinguish the consecutive packet collision introduced by the MITM attacker from normal packet collisions.
	\item We evaluate the performance of our proposed solution through both theoretical analysis and simulation. The proposed solution has 0 missed detection ratio and can achieve an arbitrarily low false positive ratio.
\end{enumerate}

The remainder of this paper is organized as follows. Related works are briefly reviewed in section II. Section III presents the adversary model and the attacks behavior model. The proposed key establishment protocol and the attacker detection mechanism is introduced in section IV. Section V and section VI show the performance of the proposed solution through theoretical analysis as well as numerical and simulation results. We discuss the impersonation attack in section VII and conclude this paper in section VIII.

\section{Related Work}

Establishing a shared secret key over a public channel can be achieved using a cryptographic method such as the DH key exchange ~\cite{diffie1976new}. However, running the standard DH key exchange protocol over a public channel is vulnerable to the MITM attack. Many research efforts have been devoted to developing device pairing protocols, which usually leverage OoB channels to provide the mutual authentication required to prevent the MITM attack. The OoB channel is assumed to possess certain security properties, for example, it is only accessible by the legitimate users, which helps verify the message source. Over the past decades, there has been many research focusing on optimizing the computation and communication cost of the pairing protocol, as well as utilizing different forms of OoB channels to improve the user-friendliness during the pairing process~\cite{balfanz2002talking,gehrmann2004manual,hoepman2004ephemeral,wong2007multichannel,vaudenay2005secure,laur2006efficient,mccune2005seeing,saxena2006secure}.

OoB channels usually require non-trivial human effort and advance user interfaces. Typical OoB channels used in device pairing includes but not limited to:  hardware port and extra cable~\cite{stajano1999resurrecting}; visual channels such as device screen~\cite{mccune2005seeing,balfanz2004network}, LED lighting~\cite{roth2008simple}; audio channels such as speaker and microphone~\cite{soriente2008hapadep}. There are new out-of-band channels emerging with the sensing modalities in advanced smart devices, for example, synchronized drawing~\cite{sethi2014commitment}, shaking together~\cite{mayrhofer2009shake}.  There are also methods which exploit the shared environmental context, such as the lightening and sound, to verify the near proximity of the involved devices\cite{miettinen2014context,schurmann2013secure}. Again, these solutions require additional sensing modalities that are not available to all devices, and cannot prevent a nearby MITM attacker that sharing the same physical context.

There are a few attempts to develop in-band solutions for initial trust establishment\cite{gollakota2011secure,hou2013chorus,shen2016secure}.  The insight of these solutions is similar to the I-Code technique\cite{vcapkun2008integrity}, which protects the integrity of the message payload by modulating the message signal with Manchester coded ON/OFF keying. Specifically, in ~\cite{gollakota2011secure}, the author designed a tamper-evident announcement (TEA) message format which improves the performance and fixed the security vulnerability of the \emph{I-codes} by introducing an exceptional long synchronization packet to guarantee an adversary cannot hide the fact that a TEA message is being transmitted. However, the solution in \cite{gollakota2011secure} relies on the push button configuration to provide authentication, which in fact is an OoB channel. \cite{hou2013chorus} proposed an in-band solution for trust establishment among multiple users, which can compare multiple authentication strings at the same time. \cite{shen2016secure} developed a group key establishment protocol for 802.15.4 based network, in which message self-authentication is achieved by combining the I-code integrity guarantee property and the transmission pre-scheduling function of the 802.15.4 superframe structure. However, it relies on the assumption that there exists a trustable coordinator. Besides, these I-Code solutions require modification to the physical layer signal modulation method.

In this paper, we propose a novel link layer solution for wireless device pairing. Compared to the existing solutions, our method works completely in-band and does not require any human interaction or special hardware. Besides, the implementation cost of our solution is minimal: no modification to the device's physical layer transmission mechanism is involved, and a user only needs to configure its backoff counter during the key establishment process.

\section{MITM attack modeling in 802.11 based network}
In this section, we will introduce our adversary model, and model the attacker's behavior for the wireless MITM attack on the message level. We will present the best strategy for the MITM attacker in an 802.11 based wireless network. We will also describe our system model and problem statement.
\subsection{MITM Attack in Diffie-Hellman Key Exchange}
\begin{figure}
	\centering
	\subfloat[The Diffie-Hellman key exchange.]{
		\label{subfig:dh}
		\includegraphics[trim={0 6cm 0 3.5cm}, clip,width=0.35\textwidth]{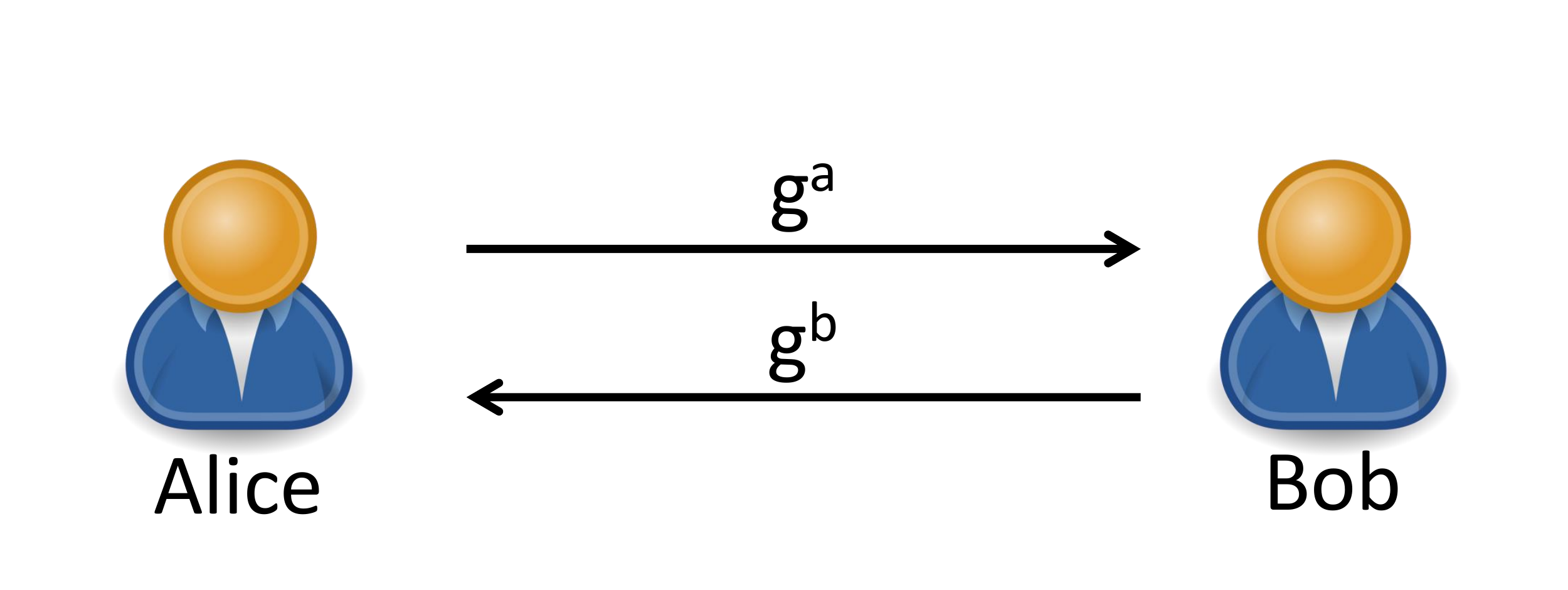} } 
	
	\subfloat[The MITM attack scenario.]{
		\label{subfig:mitm}
		\includegraphics[trim={0 6cm 0 3.5cm}, clip,width=0.35\textwidth]{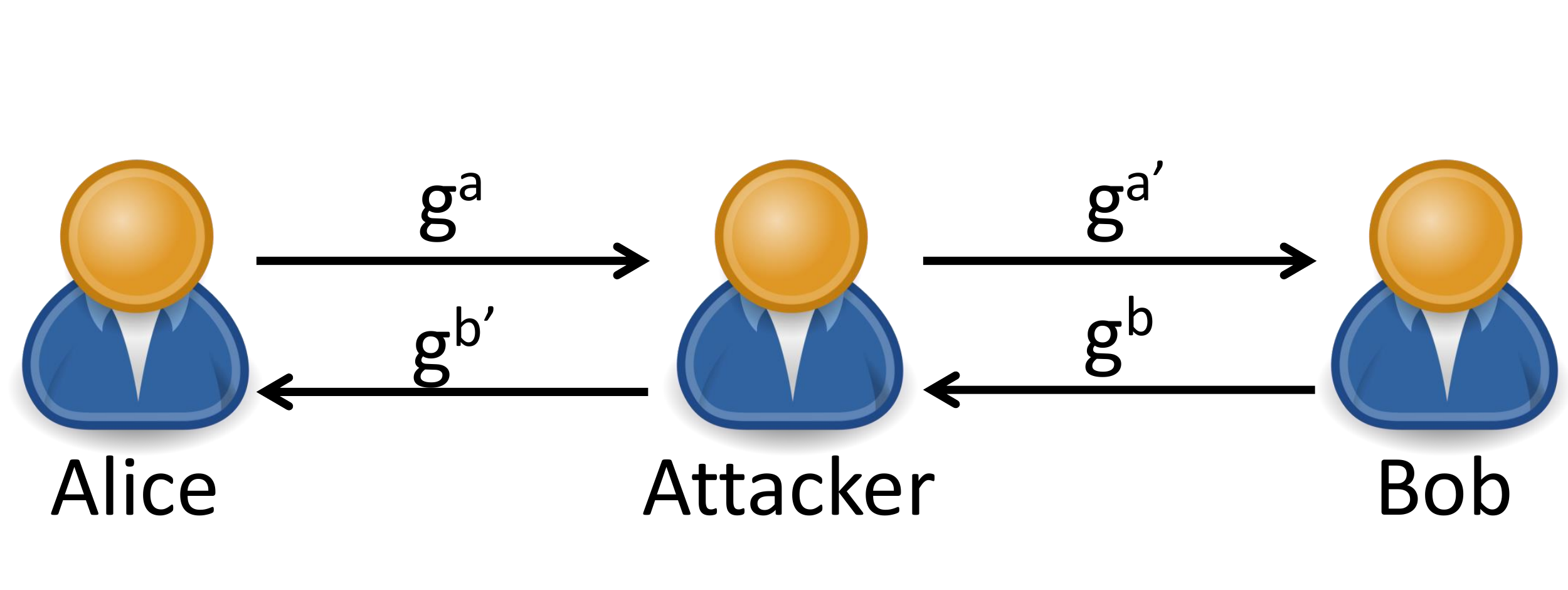} } 
	~ %add desired spacing between images, e. g. ~, \quad, \qquad, \hfill etc. 
	
	\caption{The Diffie-Hellman key exchange protocol and the MITM attack scenario.}\label{fig:dhmitm}
\end{figure}

The DH key exchange allows two parties that have no prior knowledge of each other to jointly establish a shared secret key over a public insecure channel. As shown in Fig. \ref{subfig:dh}, the DH key exchange protocol works as follows: Assume Alice and Bob agree on a large prime $p$ and a finite cyclic group $\mathcal{G}$ of order $n$ with a generator $g$. Alice randomly picks a secret value $a$ ($0\leq a \leq p-1$) and calculates $g^a\mod{p}$, and Bob randomly picks a secret random value $b$ ($0\leq b \leq p-1$)and calculates $g^b\mod{p}$. Then Alice and Bob exchange their value of $g^a$ and $g^b$ (all values are $\mod{p}$ unless otherwise specified). At the last stage, Alice calculates $K_A=(g^a)^b$ and Bob calculates $K_B=(g^b)^a$. Both Alice and Bob will arrive at the same secret value $K$ since $(g^a)^b=(g^b)^a$. 

Although the DH key exchange is secure against passive attackers (eavesdroppers), as is well know, it is vulnerable to the MITM attack. Since the message transmission is conducted over a public channel, and there is no pre-shared secret between Alice and Bob, the received $g^a$ and $g^b$ cannot be authenticated. In the MITM attack scenario, the attacker intercepts the legitimate messages and forges fake ones. For example, as shown in Fig. \ref{subfig:mitm}, the attacker intercepts the message $g^a$, and sends $g^{a'}$ to Bob pretending himself to be Alice. The attacker will also intercept $g^b$ and forge a $g^{b'}$. As a result, the attacker establishes two secret keys with Alice and Bob respectively, while Alice and Bob think they have established a shared secret key with each other. 

%\begin{figure}[htp]
%	
%	\subfloat[diffiehellman.pdf]{%
%		\includegraphics[clip,width=\columnwidth]{example-image-a}%
%	}
%	
%	\subfloat[mitm.pdf]{%
%		\includegraphics[clip,width=0.6\columnwidth]{example-image-b}%
%	}
%	
%	\caption{main caption}
%	
%\end{figure}

\subsection{Adversary Model in Wireless Communication}
The above-mentioned MITM attack is easy to be implemented in wired communications, since the attacker may get physical access to the communication cables and then intercept or manipulate the message signal without being noticed by the receiver. However, it is not the case in wireless communications. In wireless communication, message transmissions take place over the open shared wireless medium, and it is impossible for the attacker to take full control of the wireless channel. As a result, some common attacker vectors such as message interception and message modification are extremely difficult, if not impossible, to be practically implemented in wireless communications. Although designing a system with an overestimation of the attacker's capabilities does not harm its security performance, it usually makes the solutions overcomplicated by introducing unnecessary communication and computational overhead or make the hardware setup more costly. The attacker model of any communication system should base on a realistic assessment of the system vulnerabilities and attacker capabilities.

The most commonly used, which is also the strongest attacker model is the Dolev-Yao model~\cite{dolev1983security}, in which the attacker has the capability to eavesdrop, modify, compose, and replay any messages transmitted and received by legitimate devices. In this model, in addition to eavesdropping and insertion, the attacker can fully modify and annihilate signals at the receiver's antenna.  ~\cite{popper2011investigation} systematically investigated the applicability of the Dolev-Yao model in the wireless communication system. Based on their theoretical analysis and simulation, in the worst case (best case for the attacker), the attacker has a low chance (13.5\%-25\%) to deterministically flip a single bit (which is much more easier than arbitrarily manipulating an ongoing message), and can covertly annihilate a signal without being detected by existing energy-based jamming detection countermeasures. However, the worse case scenario requires the attacker to be able to measure distances and estimate the channel with high precision to any target node, and be able to achieve perfect carrier phase synchronization and precisely control the signal amplitude levels at the target receiver. In a real-world wireless network, where nodes communicate over time-varying fading channels, the state information of the sender-receiver channel is not available to the attacker, and carrier phase synchronization and amplitude control at the target receiver is not possible. {\it To the best of our knowledge, there is no existing techniques which can practically manipulate or annihilate an ongoing wireless message.} We can safely assume that in realistic settings, deterministic message modification and message annihilation is not achievable even for the strongest attacker.

In this paper, we consider a strong but realistic adversary model. We assume the attacker can eavesdrop and replay legitimate wireless messages, and he can insert messages at arbitrary times. The attacker has access to the state-of-art RF techniques such as beamforming, and can spatial selectivity transmit or jam signals to one particular target. The only limitation of the attacker is that he cannot arbitrarily manipulate an ongoing wireless message, neither can he annihilate an ongoing wireless message to the noise level. Here we do not consider that an attacker utilizes the capture effect to manipulate a message, since a significantly overpowered signal transmitted by the attacker can be detected by applying a received signal strength threshold at the legitimate users.

\subsection{Wireless MITM Attacker Behavior Modeling}
\begin{figure}[!h]
	\centerline{\includegraphics[trim={0 0 0 0}, clip, width=3in]{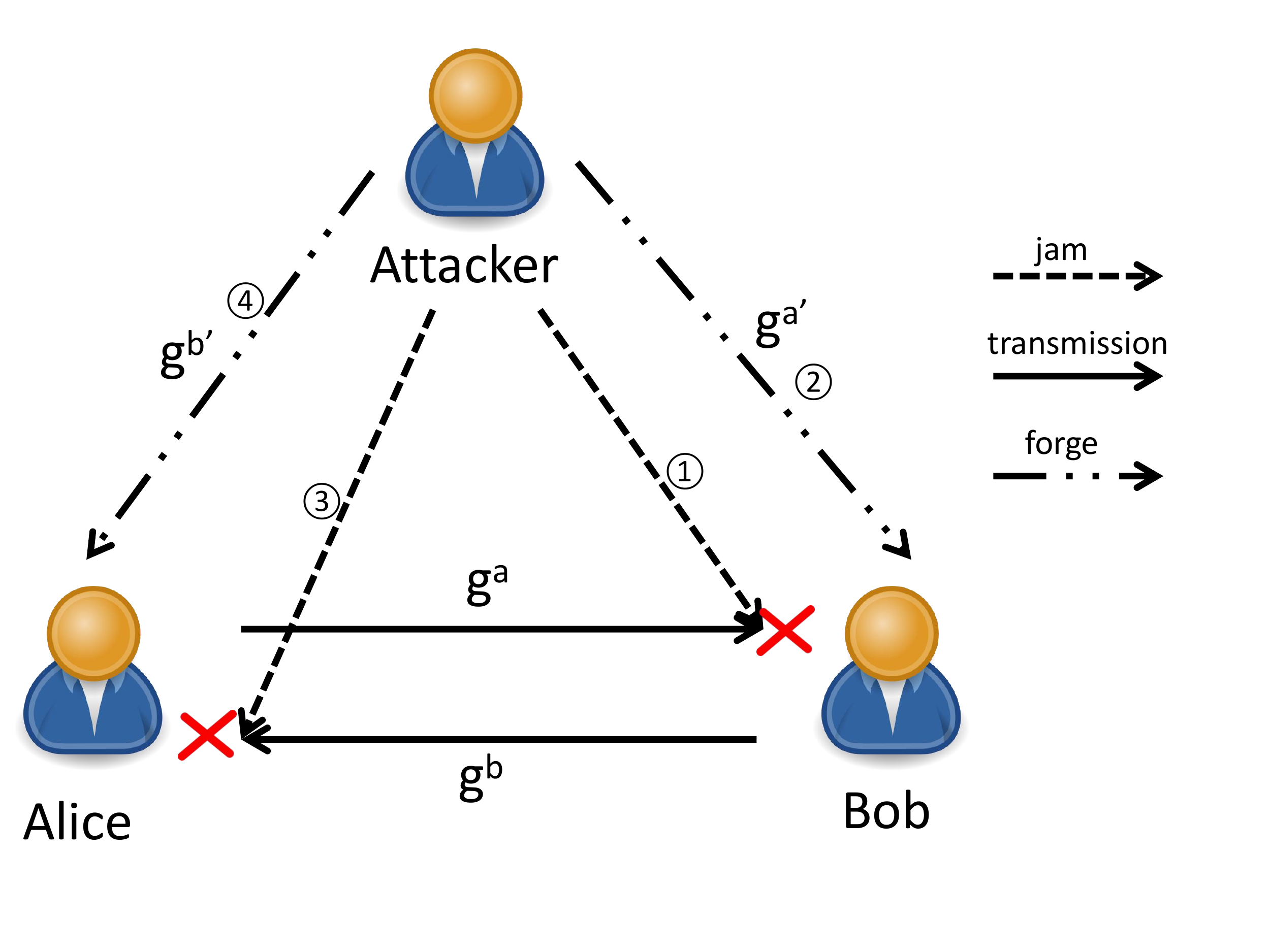}}
	\caption{Wireless MITM attack scenario.}
	\label{wirelessmitm}
\end{figure}

As we mentioned before, to successfully launch the MITM attack during the DH key exchange process, the attacker has to intercept $g^a$ and replace it with $g^{a'}$ without being noticed by Alice and Bob. Based on the previously discussed realistic attacker model, the best the attacker can do is to send a jamming signal at the same time when Alice transmits $g^a$, resulting in a packet collision at Bob's receiver, such that Bob cannot decode the message $g^a$ from Alice. This can be achieved by either jamming the complete message or jamming only the message preamble\cite{popper2011investigation}. After jamming the original $g^a$, the attacker then has to forge a $g^{a'}$ using Alice's identity criteria (usually Alice's IP address and MAC address) and sends it to Bob. At Bob's point of view, the failure of decoding the original message seems to be caused by normal packet collision, which is quite often in a random access based wireless network, such as 802.11 based and 802.15.4 based wireless networks. Bob will accept the forged $g^{a'}$ as the legitimate message from Alice since there is no authentication mechanism. The attacker then performs the same strategy to $g^b$ and successfully launched the MITM attack. The attack scenario is illustrated in Fig.  \ref{wirelessmitm}.

Based on the behavior order of the attacker, here we can further categorize the MITM attack into two categories. If the attacker follows the order of 1-2-3-4 as illustrated in Fig. \ref{wirelessmitm}, we define this type of attack as \textbf{Type I} attack. However, the attacker can also conduct the MITM attack by performing steps 1-4-2-3. In this case, the attacker first impersonates Bob to establish a shared key with Alice, then impersonates Alice to establish a shared key with Bob. In this paper, we term the second type of attack as \textbf{Type II} attack.

It is worth noting that for the transmission of the inserted messages $g^{a'}$ and $g^{b'}$, the attacker has to use directional antenna such that the party being impersonated cannot receive the message signal, otherwise the attack will be detected with minor efforts: Alice and Bob can keep monitoring the channel, and raise an alarm if they see any packet being transmitted is using their own IP and MAC addresses.  

Now let's consider a real-world application scenario, where Alice and Bob communicate via an IEEE 802.11 based WiFi network. In 802.11 based wireless networks, after successively receiving a data packet, the receiver replies to the sender with an acknowledgment frame (ACK) to indicate the correct reception. In this case, if $g^a$ is jammed by the attacker, Bob will not correctly decode this message, thus Alice will not receive the corresponding ACK. After ACK\_TIMEOUT, Alice will try to retransmit the message, until a maximum retransmission counter is reached. If the attacker keeps jamming all the retransmission from Alice, Alice will notice the abnormal behavior of the wireless channel and may stop trying to establish the secret key with Bob. If the attacker ignores Alice's retransmission of $g^{a}$ and still forges his $g^{a'}$, Bob will receive both key exchange messages, which indicates the existence of the attacker. In both cases, the MITM attack will not succeed. The best strategy for the attacker is after jamming $g^a$, he forges an ACK to make Alice believe that the message $g^a$ has been received correctly by Bob, then the attacker can forge a $g^{a'}$ and send it to Bob. However, after receiving the forged $g^{a'}$, Bob will send an ACK, and the reply address of this ACK will be Alice's MAC address. The attacker has to jam this ACK to prevent Alice from receiving double ACKs for only one message transmission. 

\begin{figure}[!ht]
	\centerline{\includegraphics[trim={1cm 1.5cm 1cm 1.5cm}, clip, width=3.2in]{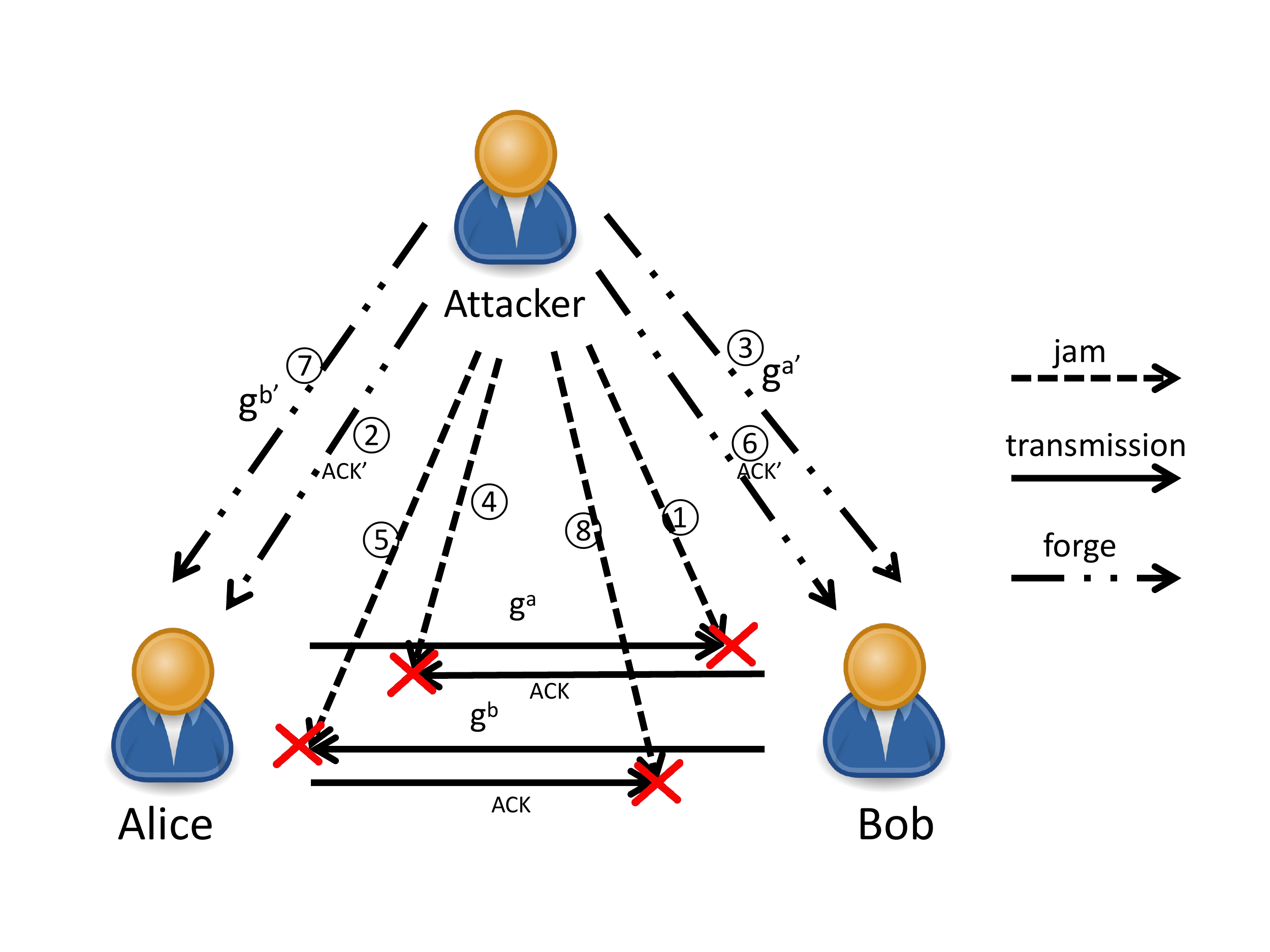}}
	\caption{MITM attack in 802.11 based network.}
	\label{80211mitm}
\end{figure}

We present the attacker's best strategy using Fig. \ref{80211mitm}. In summary, to successfully launch the MITM attack in an 802.11 based wireless network with ACK mechanism, the attacker has to follow the 8 steps illustrated in Fig. \ref{80211mitm}. A \textbf{Type I} attacker will follow the orders of 1-8 and a \textbf{Type II} attacker will follow the steps of 1-2-7-8-3-4-5-6. 

\subsection{System Model and Problem Statement}

We consider a one-hop wireless ad hoc network in which Alice and Bob communicate with each other
via an IEEE 802.11 based wireless connection. Assume that Alice and Bob have already gone through an association handshake, in which Alice sends the association request and Bob replies with an association reply to set up the wireless link between them. Immediately after the association handshake, Alice and Bob try to establish a shared secret key by a DH key exchange. Suppose Alice initializes the key exchange protocol. During the DH key exchange process, an attacker with capabilities defined in section 3.2 may try to launch the MITM attack, whose best strategy has been analyzed and illustrated in section 3.3. Besides Alice, Bob and the attacker, there are $n$ $(n\geq 0)$ other wireless stations sharing the same wireless channel. We assume that Alice and Bob have no data packets to transmit other than the messages related to the key exchange protocol, while other wireless stations have data traffic which can be treated as background traffic to the key exchange process. All the background wireless stations access the wireless channel according to the distributed coordination function (DCF) specified in IEEE 802.11 standards. We assume the channel quality is perfect, i.e., the bit error rate is 0. We further assume that  the $n$ background stations are within the transmission and receiving range of Alice and Bob (no hidden terminals), and their network traffic density is stable.  

%
%\begin{figure*}[ht]
%	\centerline{\includegraphics[trim={1cm 20.8cm 0cm 1.3cm}, clip, width=6in]{dcf.pdf}}
%	\caption{Example of DCF operation.}
%	\label{dcf}
%\end{figure*}
%

Based on the above system model and assumptions, in this paper, we aim at addressing the following problem: How can Alice and Bob prevent or detect the MITM attack during their key exchange process while using only in-band channel? We consider the attacker as being detected if both Alice and Bob detect its presence. We are seeking practical solutions that are easy to be implemented, and require no modification to existing wireless hardware. In this paper, we do not consider the denial of service (DoS) attack in which the attacker only prevents Alice and Bob from agreeing on the same key, since this can be easily achieved by keep jamming the wireless channel.

\section{An In-Band Secure Key Establishment Protocol}
\subsection{Overview of the Proposed Solution}
The insight of our solution is as follows. From the attacker's behavior analysis in section 3, we notice that in order to prevent Alice or Bob receiving legitimate key exchange messages, the attacker has to transmit jamming signals to intentionally collide the legitimate messages at the receiver's antenna. This deterministic packet collision introduced by the attacker can be observed by the receiver. To distinguish the packet collision introduced by the MITM attacker from normal packet collisions due to simultaneous transmission, our protocol requires Alice to transmit $g^a$ multiple times consecutively, such that to successfully launch the MITM attack, the attacker has to jam all the $g^a$ from Alice, thus resulting in a burst sequence of packet collisions at Bob's receiver. Then Bob can notice this abnormal channel behavior and detect the existence of the attacker. We will achieve this by modifying the DH key exchange protocol and the channel access scheme for the protocol message transmission, as well as carefully designing an attacker detection mechanism. 

\subsection{Preliminary: 802.11 Distributed Coordination Function (DCF)}
In this subsection, we briefly describe the main procedures in the DCF of 802.11 MAC protocol, which is a carrier sense multiple access with collision avoidance (CSMA/CA) mechanism. In the 802.11 MAC protocol, a station with a packet to transmit should first monitor the channel before transmitting. If the channel is sensed to be idle for a period of time equals to a distributed interframe space (DIFS), then the station selects a random backoff counter. The backoff counter is uniformly selected from $[0, CW]$, with $CW$ being its current contention window size. The station decreases its backoff counter by 1 for each time slot when the channel is sensed idle after the DIFS. If the channel is sensed to be busy during the backoff procedure, the station will freeze its backoff counter until the end of current transmission, and reactivate the backoff counter until the channel is sensed idle again for more than a DIFS time period. The station transmits when its backoff counter reaches 0. A packet collision happens when more than one stations have their backoff counter reaching zero at the same time slot and transmit simultaneously. 

In the CSMA/CA mechanism, a station does not have the capability to detect the collision of its own transmission, so an ACK is transmitted by the receiver station to indicate the successful packet reception. The ACK is transmitted after a period of time equals to a short interframe space (SIFS) immediately follows the successful reception. Since the SIFS is shorter than the DIFS, no other stations are able to transmit until the transmission of the current ACK finishes. In other words, the transmission of the ACK always has a higher priority compared to normal data packet transmission. If a station does not receive an ACK within a time period of ACK\_TIMEOUT after it transmits a packet, it knows that its previous transmission has failed. Each time a transmission fails, the contention window size of this station $CW$ doubles its value before it reaches a maximum upper limit $CW_{max}=2^{\beta}CW_{min}$, and remains to be this value until the maximum retransmission limit $\alpha$ is reached. When the maximum retransmission limit is reached, $CW$ will be reset to $CW_{min}$ and the current data packet will be discarded.

\subsection{The Proposed Key Exchange Protocol}
\begin{figure}[!ht]
	\centerline{\includegraphics[trim={0 0 2cm 0}, clip, width=3.3in]{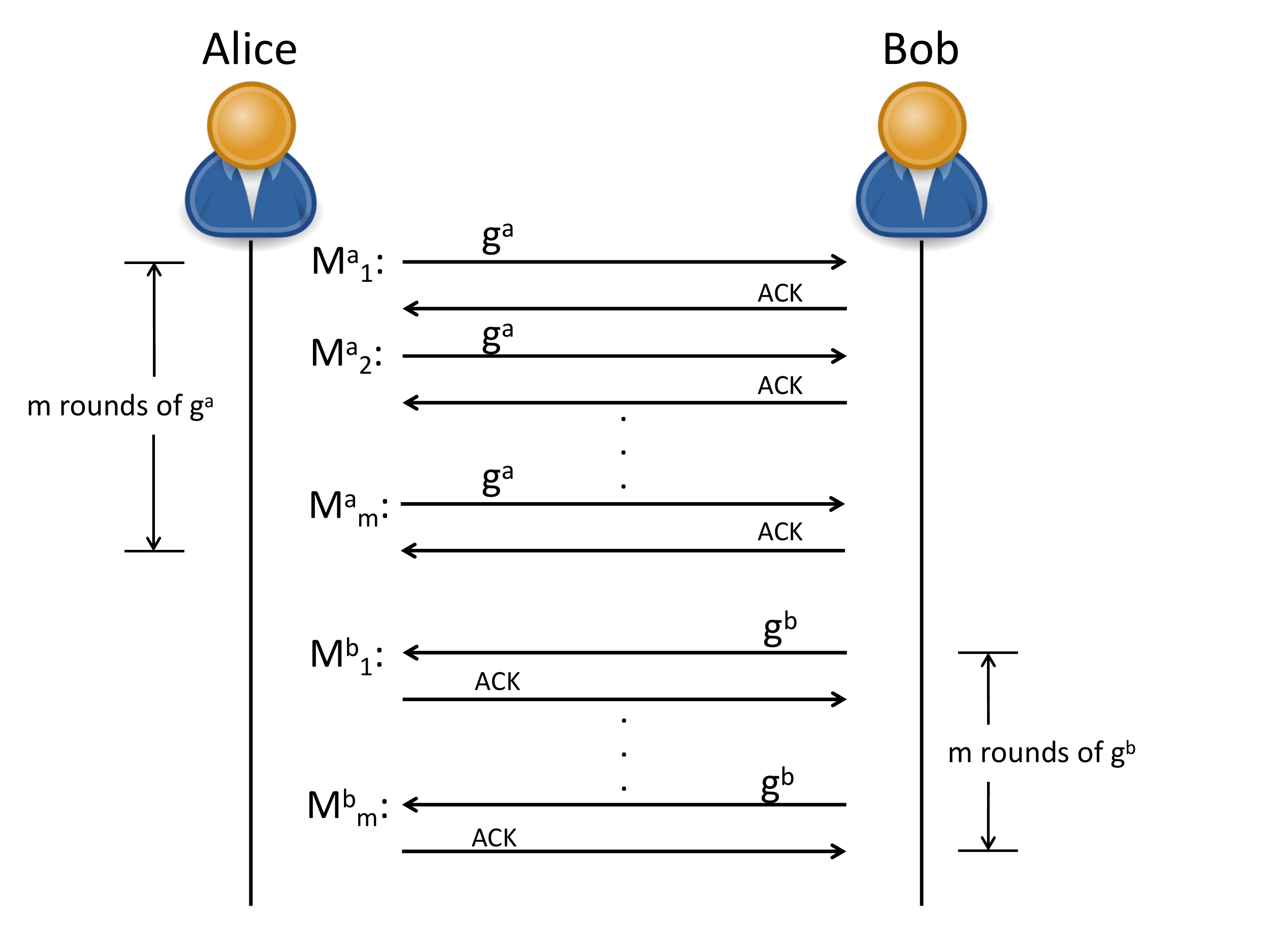}}
	\caption{The proposed key exchange protocol.}
	\label{fig:protocol}
\end{figure}

As we mentioned, the insight of our solution is to let Alice and Bob transmit the DH key exchange message multiple times. In order to successfully launch the MITM attack, the attacker has to jam all these parameters at the receiver's antenna, resulting in a burst sequence of packet collisions such that the receiver can distinguish this abnormal channel behavior from normal packet collision. The modified DH key exchange protocol is illustrated in Fig. \ref{fig:protocol}. We use $M^a_{i}$ and $M^b_{i}$ denote the $i$th message from Alice and Bob, respectively; the superscript is dropped when we generally indicate a message from either Alice or Bob.

In the modified DH key exchange protocol, each key exchange message is transmitted $m$ times. In this situation, to successfully perform the MITM attack, the attacker has to block all these $2m$ messages by introducing $2m$ extra packet collisions to the channel, and Alice and Bob can each observe $m$ packet collisions. The value of $m$ shall be determined based on the current channel condition, which will be discussed in later sections of this paper. 

We set the message format for $M_i$ as follows:
\begin{equation*}
	M_i = \{\;i\;|\;\;m\;|\;g^a\,(\,\text{or}\;g^b\,)\;|\;\text{dummy data}\},
\end{equation*}
where $i$ denotes the current message number and $m$ denotes the total number of messages (rounds) in the key exchange protocol. $g^a$ or $g^b$ is the same as the original DH protocol. We use dummy data to fill the message payload to be the maximum size allowed in IEEE 802.11 standard (typical value being 2304 Bytes). The reason we use dummy data to reach the maximum frame size is to prevent the attacker jamming more than one packets by sending one exceptionally long jamming signal. For example, if $m=2$, and the length of each $M_i$ is only 500 Bytes, our goal is to let the receiver observe 2 packet collisions when an attacker presents. In this scenario, if the attacker uses one single jamming signal whose duration can cover part of the transmission of $M_1$ plus at least the packet header of $M_2$, the receiver will only observe one collision event. If we set the packet size of the key exchange protocol to be the maximum size allowed in the 802.11 standards, when the receiver observes a collision whose duration is longer than the transmission time of a maximum-sized packet, it will notice this abnormal channel behavior. We term a packet collision with a duration longer than the time required to transmit a maximum-sized packet as an exceptionally long packet collision.
%The message format for the proposed protocol is shown in Fig. \ref{frame}. 
%\begin{figure}[!ht]
%	\centerline{\includegraphics[trim={0.5cm 21.5cm 6.5cm 3cm}, clip, width=3.5in]{frame.pdf}}
%	\caption{Message format.}
%	\label{frame}
%\end{figure}
%
%\begin{figure}
%	\centering
%	\begin{subfigure}[c]{0.45\textwidth}
%		\includegraphics[trim={1cm 21.5cm 4.5cm 2.2cm}, clip, width=\textwidth]{maxsize1.pdf}
%		\caption{small packet size}
%		\label{fig:maxsize1}
%		\vspace{0.5cm}
%	\end{subfigure}
%	~ %add desired spacing between images, e. g. ~, \quad, \qquad, \hfill etc. 
%	%(or a blank line to force the subfigure onto a new line)
%	\vspace{0.5cm}
%	\begin{subfigure}[c]{0.45\textwidth}
%		\includegraphics[trim={1cm 21cm 4.5cm 2.2cm}, clip, width=\textwidth]{maxsize2.pdf}
%		\caption{max packet size}
%		\label{fig:maxsize2}
%	\end{subfigure}
%	
%	\begin{subfigure}[c]{0.45\textwidth}
%		\includegraphics[trim={1cm 21cm 4.5cm 2.2cm}, clip, width=\textwidth]{maxsize3.pdf}
%		\caption{max packet size}
%		\label{fig:maxsize3}
%	\end{subfigure}
%	~ %add desired spacing between images, e. g. ~, \quad, \qquad, \hfill etc. 
%	%(or a blank line to force the subfigure onto a new line)
%	
%	\caption{Use maximum packet size for protocol message transmission.}\label{fig:maxsize}
%	\label{fig:maxsize}
%\end{figure}
\begin{figure}
	\centering
	\subfloat[Small packet size.]{
		\label{subfig:maxsize1}
		\includegraphics[trim={1cm 21.5cm 4.5cm 2.2cm}, clip, width=0.37\textwidth]{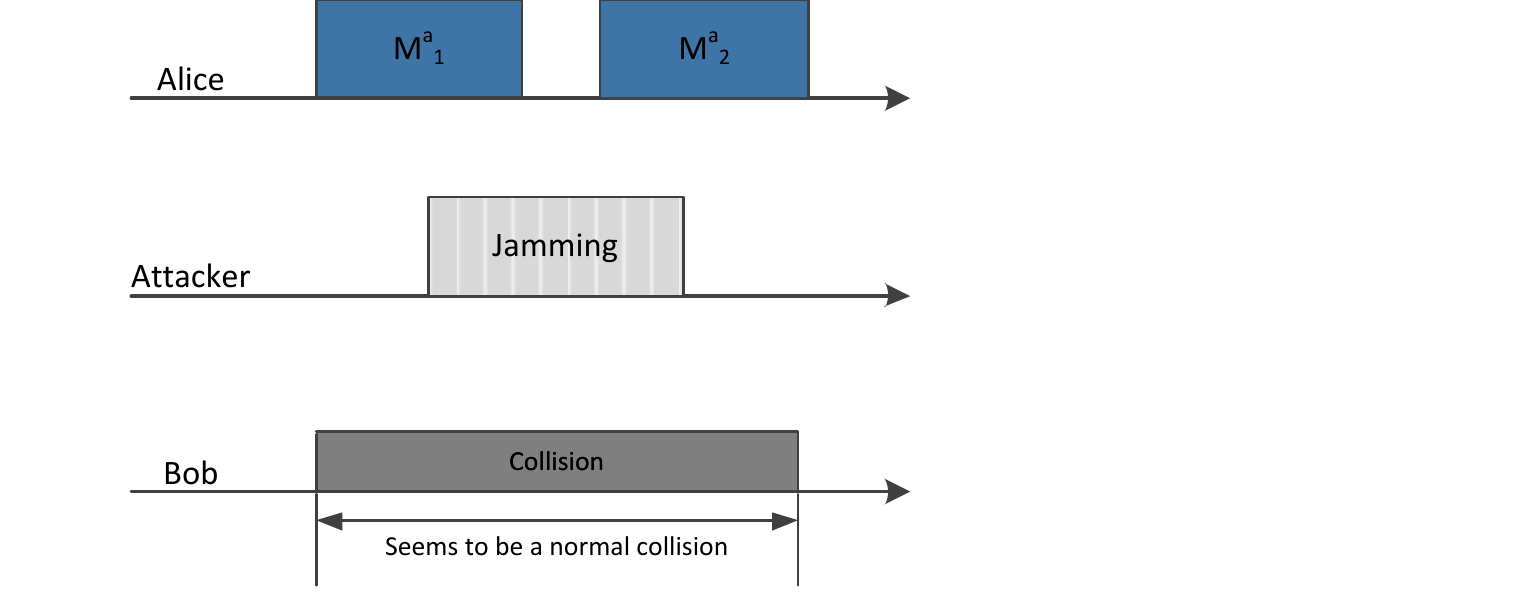} } 
	
	\subfloat[The attacker uses one jamming signal to jam two packets.]{
		\label{subfig:maxsize2}
		\includegraphics[trim={1cm 21cm 4.5cm 2.2cm}, clip, width=0.37\textwidth]{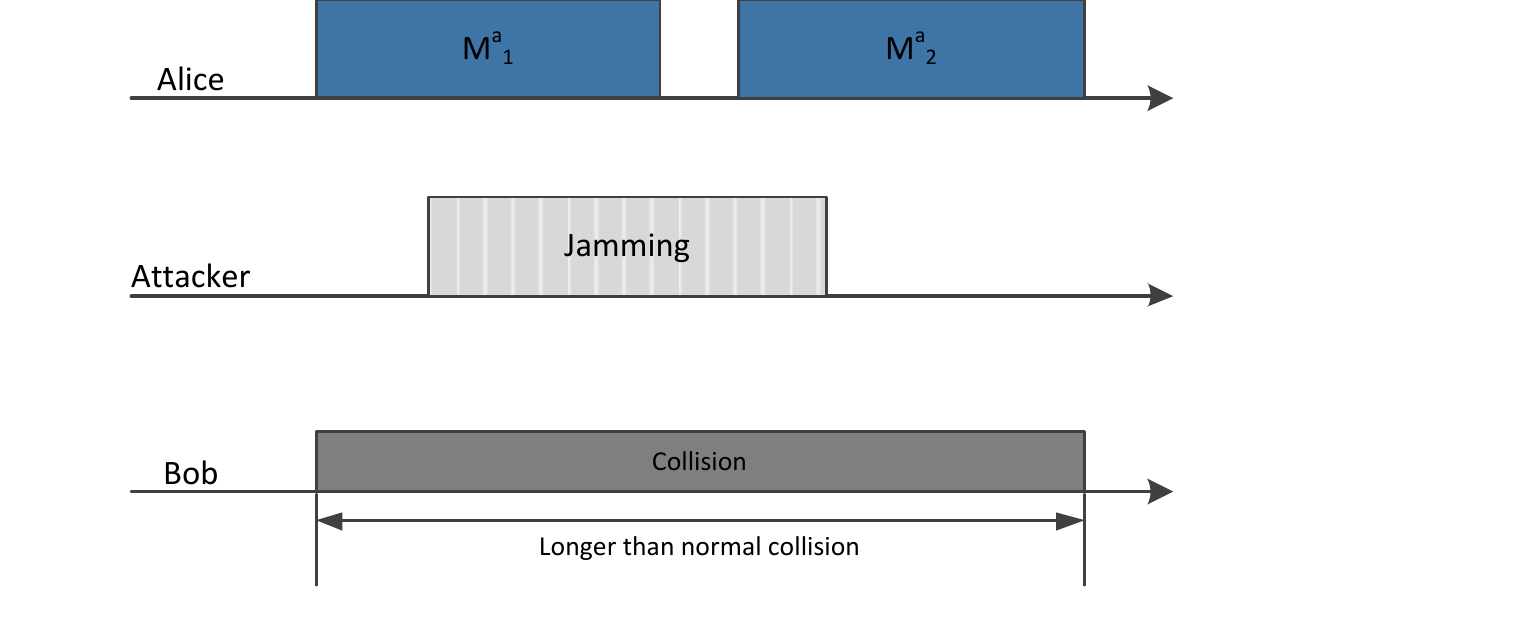} } 
	
	\subfloat[The attakcer uses two jamming signals.]{
		\label{subfig:maxsize3}
		\includegraphics[trim={1cm 21cm 4.5cm 2.2cm}, clip, width=0.37\textwidth]{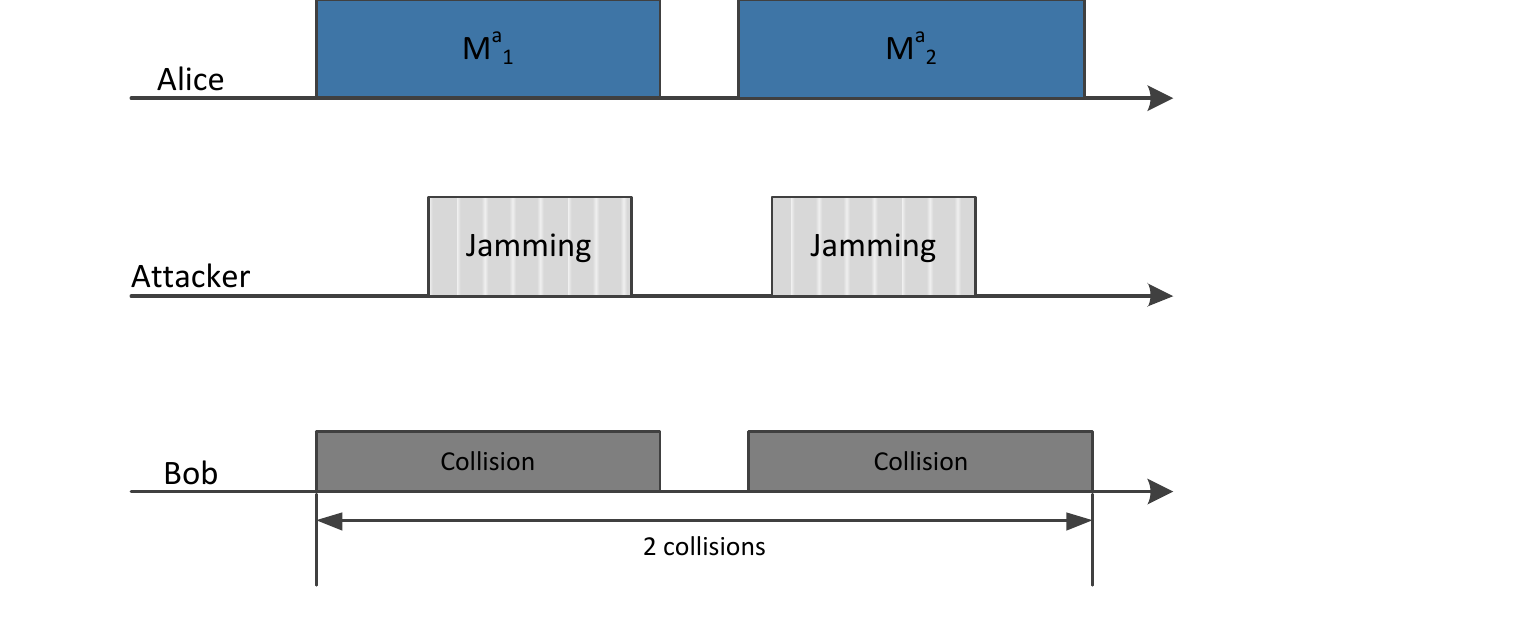} } 
	~ %add desired spacing between images, e. g. ~, \quad, \qquad, \hfill etc. 
	
	\caption{Use maximum packet size for protocol message transmission.}\label{fig:maxsize}
\end{figure}

Fig. \ref{fig:maxsize} illustrates the reason why we use dummy data to fill up the message payload to reach the maximum packet size allowed. In Fig. \ref{subfig:maxsize1}, Alice transmits $g^a$ two times without adding dummy data, and the attacker jams the message with one jamming signal. In this case Bob will detect one collision event. If the two packet transmission duration plus their inter-frame space is less than the transmission time of one maximum-sized packet, this will seem to be one normal packet collision for Bob. In Fig. \ref{subfig:maxsize2}, Alice transmits $g^a$ with dummy data added, and the attacker tries to jam the message transmission with only one jamming signal. In this case, Bob will be able to observe a collision longer than normal packet transmission, thus detect the existence of the attacker. If the attacker wants to jam the maximum-sized key exchange message without being detected by the collision duration criteria, he has to jam each message separately, as shown in Fig. \ref{subfig:maxsize3}. In summary, with dummy data added to the key exchange messages, if the attacker attempts to prevent Bob from receiving Alice's key exchange messages, Bob is guaranteed to observe $m$ packet collisions. The same is true for the key exchange messages from Bob to Alice.

\begin{figure*}[ht]
	\centering
	\subfloat[Channel behavior without the attacker.]{
		\label{subfig:access1}
		\includegraphics[trim={1cm 23.5cm 15.5cm 0.9cm}, clip, width=0.75\textwidth]{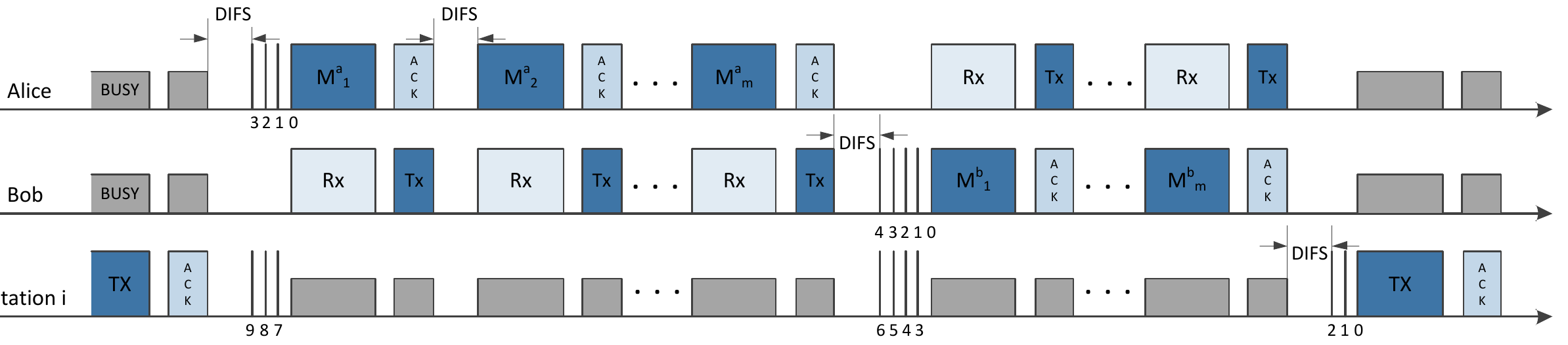} } 
	
	\subfloat[Channel behavior in presence of the attacker.]{
		\label{subfig:access2}
		\includegraphics[trim={1cm 22cm 12.5cm 0.8cm}, clip, width=0.75\textwidth]{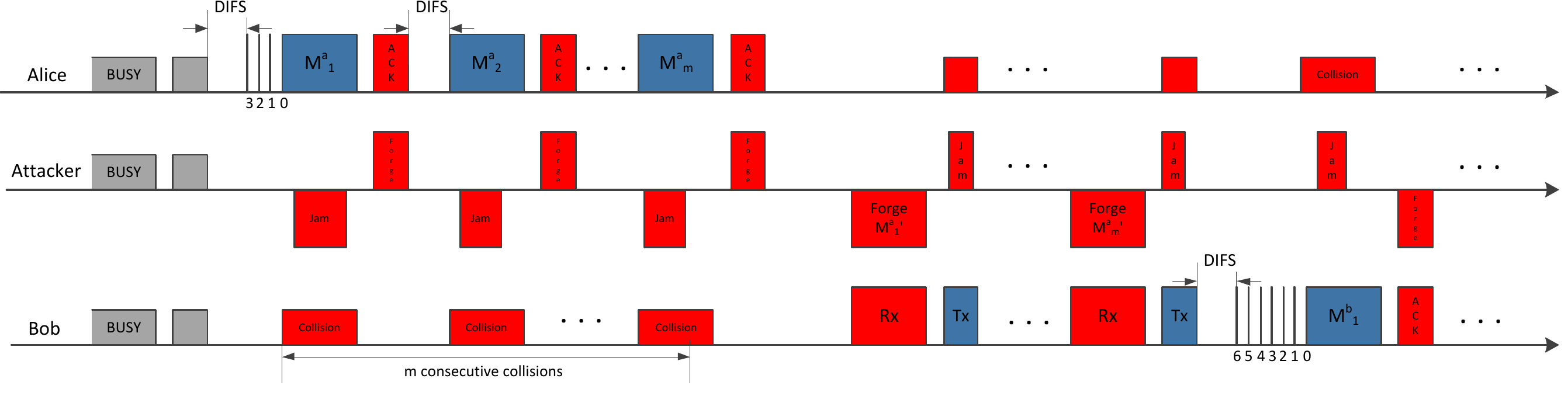} }

	~ %add desired spacing between images, e. g. ~, \quad, \qquad, \hfill etc. 
	
	\caption{Channel behavior of the proposed scheme.}\label{fig:access}
\end{figure*}

Our proposed key exchange protocol guarantees that if the attacker wants to successfully launch the MITM attack, both Alice and Bob will be able to observe $m$ extra packet collisions. The remaining problem is how to distinguish these extra packet collisions introduced by the attacker from normal ones caused by simultaneous transmissions, since packet collision frequently happens due to the distributed nature of the channel access mechanism: if the backoff counters of two or more stations happen to reach 0 at the same time slot, a packet collision will occur. In fact, every station within the wireless network shares equal chance to access the wireless channel. Suppose Alice and Bob share the channel with $n$ other stations, then on average Alice has to wait for $n$ packet transmission before it gets a chance to transmit a single protocol message. This implies that in presence of the attacker, on average Bob can only observe one extra collision for every $n+1$ packet transmissions, which cannot serve as a good detection criterion.

%\subsection{Modified Channel Access Scheme for the Proposed Key Exchange Protocol}
%\begin{figure*}[ht]
%	\centering
%	\begin{subfigure}[c]{0.9\textwidth}
%		\includegraphics[trim={1cm 23.5cm 15.5cm 0.9cm}, clip, width=\textwidth]{access1.pdf}
%		\caption{Channel behavior without the attacker}
%		\label{fig:access1}
%		\vspace{0.5cm}
%	\end{subfigure}
%	~ %add desired spacing between images, e. g. ~, \quad, \qquad, \hfill etc. 
%	%(or a blank line to force the subfigure onto a new line)
%	\vspace{0.5cm}
%	\begin{subfigure}[c]{0.9\textwidth}
%		\includegraphics[trim={1cm 22cm 12.5cm 0.8cm}, clip, width=\textwidth]{access2.pdf}
%		\caption{Channel behavior in presence of the attacker}
%		\label{fig:access2}
%	\end{subfigure}
%	
%	
%	\caption{Channel behavior of the proposed scheme.}\label{fig:maxsize}
%	\label{fig:access}
%\end{figure*}

Our goal is to let Alice or Bob distinguish the extra packet collisions introduced by the attacker from normal packet ones. To achieve this, we can grant the key exchange protocol packets a higher priority to access the wireless channel compared to regular data packets, such that the extra packet collisions introduced by the attacker will not be diluted by normal collisions.

As we introduced in section 4.2, the backoff counter of a station only starts to decrease after the channel is idle for at least DIFS, and a station can only transmit if its backoff timer reaches 0. If we let the key exchange messages access the channel without going through the backoff process by fixing the backoff counter to be 0, message $M^a_i$ and $M^b_i$ will have priority over the background data packets.

The channel access scheme for the proposed key exchange protocol is as follows. Upon receiving the ACK of $M^a_{(i-1)}$, Alice will transmit $M^a_i$ ($2\leq i\leq m$) immediately after a DIFS, and upon receiving the ACK of $M^b_{(i-1)}$, Bob will transmit $M^b_i$ ($2\leq i\leq m$) immediately after an DIFS. The channel access scheme for $M^a_1$ and $M^b_1$ follow the normal backoff mechanism defined in the 802.11 standards. 
%	\item if there are background data packets (number of background wireless users $n \geq 1$), Alice will monitor the channel and wait until an ACK of the background packet has been transmitted, then he will transmit $Ma_1$ immediately after SIFS. By doing this, the transmission of $Ma_1$ is guaranteed to not collide with any background data packet or ACK.
%	\item if there is no background data packets ($n=0$, Alice and Bob are the only users of the wireless channel), Alice can immediately transmit $Ma_1$. The presence of background data packets can be decided by monitoring the channel for a certain period of time (for example, 100$\times$DIFS). If there is no packets transmitted during this time period, Alice believes that the channel is contention free and there is no other data packets. Here we ignore the probability that a background station happens to reach backoff timer 0 at the same time Alice decides to transmit.
%\end{enumerate}

This channel access scheme for the proposed key exchange protocol guarantees the channel access priority for message $M^a_i$ and $M^b_i$ ($2\leq i\leq m$). Only $M^a_1$ and $M^b_1$ compete for the channel with the other stations. There is a probability that the transmission of $M^a_1$ and $M^b_1$ encounters a collision, but the other messages in the key exchange protocol are guaranteed to be collision-free.  More importantly, this channel access scheme forces the MITM attacker to consecutively collide multiple packets, such that the receiver can distinguish these $m$ consecutive packet collisions from normal ones, thus detecting the presence of the attacker. Fig. \ref{fig:access} illustrates the channel behavior of the proposed key exchange protocol with this priority channel access scheme.

Fig. \ref{subfig:access1} shows the channel behavior of the proposed key exchange protocol under normal conditions (without the MITM attacker). Besides Alice and Bob, there are other stations trying to access the wireless channel. Alice starts the key exchange protocol by transmitting $M^a_1$ after winning the channel competition. Other stations (station $i$ as an example) can only get access to compete for the channel after the transmission of $M^a_m$ finishes, because $M^a_i$ ($2\leq i \leq m$) is transmitted immediately after a DIFS, and they have no chance to decrease their backoff counters. Fig. \ref{subfig:access2} illustrates the channel behavior of the proposed key exchange protocol when a \textbf{Type II} attacker presents. The figure only shows the first half of the attack, where the attacker jams all the $M^a_i$ from Alice and then forges ${M^{a'}_i}$ to Bob. The attacker's behavior results in $m$ consecutive packet collisions at Bob's receiver, which only happens with extremely low probability under normal conditions. The attacker can be then detected by our detection mechanisms (which will be introduced later). The second half of the attack (not included in Fig. \ref{subfig:access2} due to space limitations) will be the attacker jamming the $M^b_i$ and forging $M^{b'}_i$ to Alice, which will result in $m$ consecutive packet collisions at Alice's receiver. 

\subsection{System Design}
\subsubsection{Design Principles}
Given the proposed key exchange protocol and the proposed channel access scheme, the MITM attack detection algorithm design is fairly simple. The key insight for the detection algorithm design can be summarized into the following two points:
\begin{enumerate}
	\item if the attacker jams all the key exchange messages from Alice (or Bob), then Bob (or Alice) will observe $m$ consecutive packet collisions.
	\item if the attacker fails to jam all the key exchange messages from Alice (or Bob), and still try to forge an $M^{a'}_i$ (or $M^{b'}_i$), then Bob (or Alice) will receive key exchange messages containing different parameters $g^a$ and $g^{a'}$ (or $g^b$ and $g^{b'}$). 
\end{enumerate}

Base on these insights, we design our system flowchart as illustrated in Fig. \ref{fig:detector}. Before we get into the details, let's first introduce in practice how does one station determine whether an ongoing transmission is a collision.
\subsubsection{Detecting a packet collision}
In 802.11 wireless network, a station keeps monitoring the channel. Whenever the station's antenna detects an ongoing packet transmission, it will first decode the physical herder and the MAC header, and obtain the Destination Address (DA) which is contained in the MAC header. If the DA does not match its own MAC address, the station will drop the packet without attempting to decode the data payload. If the station is the designated receiver of this packet, it will continue decoding the packet data payload. If the packet can be decoded and passes the FCS check (error-detecting code), the station will send an ACK to indicate the correct reception of the packet. If the packet header or the packet payload cannot be decoded or the FCS check fails, the station will consider this packet has collided and no ACK will be transmitted.

Theoretically, we can have Alice and Bob attempting to decode all the received packets, no matter whether the DA matches or not. This way Alice and Bob can detect the packet collisions within their antenna's receiving range. However, this method will impose huge energy consumption and shorten the lifetime of energy-restricted wireless devices. Besides, if the attacker jams the preamble of a message, the receiver will not even notice that there is an ongoing message (as synchronization will fail). In this paper, we will adopt a channel collision detection method which considers the preamble jamming and is an energy-efficient. We notice that from an observer's point of view, in case of a successful packet transmission happens, the channel occupancy follows the pattern
\begin{align*}
	\text{Busy with duration}>\text{ACK}&\\
	\Longrightarrow\text{Idle}\text{ with}&\text{ duration}=\text{SIFS}\\
	&\Longrightarrow\text{Busy with duration}=\text{ACK,}
\end{align*} 
and in case of a packet collision, the channel occupancy follows the pattern
\begin{align*}
	\text{Busy with duration}>\text{ACK}\Longrightarrow\text{Idle}\text{ with}\text{ duration}>\text{SIFS.}
\end{align*} 
In our proposed scheme, Alice and Bob will use these channel occupancy patterns to determine whether an ongoing packet transmission is a successful one or a collision. 

%Let $I_n$ be the indicator denoting the state of the $n$th observed packet transmission. Specifically,
%\begin{equation*}
%I_n=\begin{cases}
% \;1,\qquad \text{the }n\text{th packet is a collision,} \\\; 0,\qquad \text{the }n\text{th packet is a successful transmission.}
%\end{cases}
%\end{equation*}
%We use $X_n$ to denote the state of the detector, then the behavior of the detector can be mathematically described as
%\begin{align}
%\begin{cases}
%X_{n+1}=I_n\times (X_n+I_n)\\
%X_0=0.
%\end{cases}
%\label{detector}
%\end{align}
%With $m$ being the detection threshold, the decision rule of the detector in step $n$ is
%\begin{equation*}
%\delta_n=\begin{cases}
%\;1 \quad \text{if} \quad X_n \geq m\\
%\;0 \quad \text{if} \quad X_n < m,
%\end{cases}
%\end{equation*}
%where $\delta_n$ is an indicator of whether $m$ consecutive collisions happen or not. The detector value $X_n$ will be reset to 0 as soon as it exceeds the threshold $m$ and the detection procedure starts over again.
\subsubsection{Flowchart of the proposed system}
\begin{figure}[!ht]
	\centerline{\includegraphics[trim={6.5cm 15.5cm 5.5cm 0.5cm}, clip, width=3.7in]{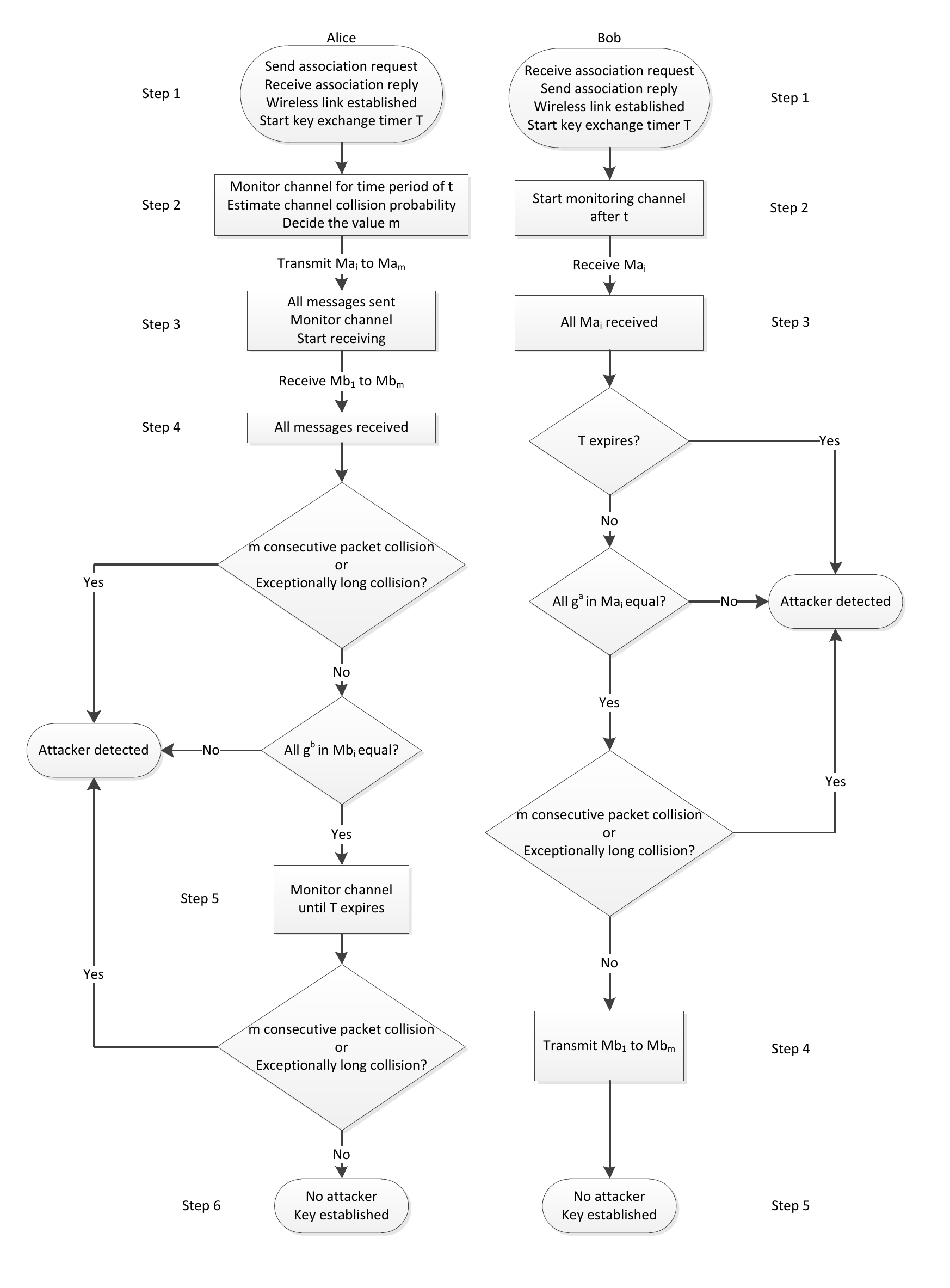}}
	\caption{System flowchart.}
	\label{fig:detector}
\end{figure}

Now let's go through the details of the proposed system. The very first step will be Alice and Bob establish their wireless link through an association handshake. Assume Alice is always the one who initiates the association handshake and the key exchange process. The system flowchart for Alice and Bob are slightly different. Lets first introduce the system process for Alice, which is shown by the left part of Fig. \ref{fig:detector}.

After receiving the association request, Alice starts a key exchange timer $T$ (step 1). Alice will only install the established key after $T$ expires and no attacker detected. Then Alice monitors the channel for a time period of $t$ before starting to transmit $M^a_1$. This time period of $t$ is termed as the monitoring window. During the monitoring window, Alice estimates the channel condition (channel collision probability and channel traffic density) and decides the number of rounds $m$ for the proposed key exchange protocol (step 2). The details of how to chose a proper $m$ will be discussed in the next section. With $m$ being decided, Alice can start the key exchange protocol by transmitting $M^a_1$ to $M^a_m$ according to the proposed channel access scheme. After finishing the transmission, Alice starts to monitor the channel and receiving $M^b_i$ (step 3). The attacker detection decision making will begin after Alice received all the $M^b_i$ (step 4). She will check if there are $m$ consecutive packet collisions between step 3 and step 4, if there are exceptionally long packet collisions, and whether the received $g^b$ are all equal. A \textbf{Type I} attacker will be detected at this point. If no attacker detected at this point, Alice still needs to monitor the channel until the key exchange timer $T$ expires (step 5) to decide whether a $Type II$ attacker exists. After $T$ expires, if no $m$ consecutive packet collision and no exceptionally long packet collision has been detected, Alice can confirm that there is no MITM attack during the key exchange process and install the established key. 

Now we describe the system flowchart for Bob. Upon receiving the association request from Alice, Bob transmits the association reply and starts his key exchange timer $T$ (step 1). Bob will start to monitor the channel after a time period of $t$ (step 2), since during this time Alice is estimating the channel condition and has not started the key exchange protocol yet. After receiving all $M^a_i$ from Alice (step 3), Bob checks if the timer $T$ expires, if all the received $g^a$ are equal, and if there are $m$ consecutive packet collisions or any exceptionally long packet collision. Both the \textbf{Type I} attacker and the \textbf{Type II} attacker will be detected by these detection criteria. If there is no attacker being detected, Bob can start transmitting his $M^b_i$. After Bob finishes the transmission of $M^b_m$, he can install the established key.
\subsection{Alternative Designs}
\subsubsection{Channel access mechanism} In the proposed solution, the protocol messages are transmitted immediately after a DIFS, without going through the back off process. This channel access mechanism aims at granting Alice or Bob continuously channel access during the protocol message transmission. In fact, this can be achieved by setting the inter-message space to be any value between SIFS and DIFS. We choose DIFS in our design for the purpose of easy implementation: Alice or Bob only need to change the backoff counter configuration in the key establishment stage.
\subsubsection{Detector design}
Given the channel access mechanism, the channel occupancy of the protocol messages has a unique pattern: the ACK of the previous message is followed by an idle period of DIFS and then a maximum-sized packet transmission. Correspondingly, if the attacker jams all the key exchange messages, the receiver will observe $m$ consecutive collisions with a fixed idle interval which equals to SIFS+ACK+DIFS. Based on this collision pattern, we can design a more advanced detector to distinguish this attacker behavior from normal packet collisions. Specifically, a timer can be added on top of the current detector design to record the timestamp for recent collisions, and if consecutive collisions are detected, the users can compare the timestamps to check if the idle intervals match the unique pattern. Compared to the current design, this more advanced detector can achieve 0 false positive ratio with the extra cost of memory space for recording the collision timestamps and extra codes to check the collision pattern. We we suggest the current solution in this paper for better performance and cost trade-off. In next section, we will show that the current detector design is sufficient and can achieve an arbitrarily low false positive ratio.     
\subsubsection{RTS/CTS mode}
The details of our solution are designed for the basic access mode in 802.11 wireless network. Besides the basic access mode, there is an RTS/CTS mode aiming at addressing the hidden terminal problem, in which each station goes through a request-to-send (RTS) and clear-to-send (CTS) handshake before the data packet transmission\cite{bianchi2000performance}. The proposed solution can be further extended to cover the RTS/CTS mode. In the RTS/CTS mode, only the RTS packet has a chance to collide with other RTS packets, while the data packet is collision-free. If the attacker only collides the data packet, the detection becomes trivial, so a successful attacker has to jam the CTS request. Base on this insight, we can have Alice transmit the RTS request $m$ times consecutively using the proposed channel access mechanism, such that the attacker has to jam all the $m$ RTS to cause $m$ consecutive collisions, which can then be detected by Bob.

\section{Performance Analysis of the Proposed Solution}
In this section, we conduct theoretical performance analysis of the proposed system in terms of missed detection ratio, false positive ratio and the cost of applying our solution. The proposed system has 0 missed detection ratio, and can achieve an arbitrarily low false positive ratio by proper parameter configuration.
\subsection{Missed Detection Ratio}
The missed detection ratio is defined as the probability that an attacker successfully launched the MITM attack without being detected by Alice or Bob. With the proposed scheme, a MITM attacker will always be detected. The missed detection ratio of the proposed scheme is 0.

The MITM attack is considered to be successful, only if the attacker can establish two separate shared keys with Alice and Bob, while passing all the detection criteria in the proposed scheme. The proposed scheme has three detection rules: (1) all received $g^a$ ($g^b$) are equal; (2) no $m$ consecutive packet collision is detected; (3) no exceptionally long packet collision is detected. In the proposed scheme, Alice will transmit $g^a$ $m$ times. If the attacker does not intercept all these $m$ messages, Bob will at least receive one $g^a$ from Alice. In this case, if the attacker transmits his own $g^{a'}\neq g^a$ to Bob, according to detection rule (1), Bob will detect his presence. The only successful chance for the attacker is that he manages to intercept all the $g^a$ from Alice. Under the adversary model, the attacker can only achieve this by colliding these messages at Bob's antenna. If the attacker uses one jamming signal to collide multiple $M^a_i$, an exceptionally long packet collision will be observed by Bob, as illustrated in Fig. \ref{fig:maxsize}, which violates detection rule (3). The attacker has to individually collide these $m$ packets. According to the proposed channel access mechanism, these $m$ packets will be transmitted consecutively without being interrupted by normal background data packets, so colliding them will result in $m$ consecutive packet collisions at Bob's receiver, which violates detection rule (2). In summary, under the adversary model defined in section 3.2, a MITM attacker can always be detected by our detection rules. The missed detection ratio of the proposed scheme is 0.

\subsection{False Positive Ratio}
A false positive occurs when the proposed system detects an attacker but in fact no threat exists. Our system will raise an alarm if any of the following situations occurs: (1) the received $g^a$ ($g^b$) does not match; (2) Alice or Bob detects $m$ consecutive packet collisions within the detection window (the time period starts after the monitoring window until the key exchange timer $T$ expires); (3) an exceptionally long packet collision is detected. Given the assumption that the channel quality is perfect (bit error equals 0) and there are no hidden terminals, if there is no attacker existing, situation (1) and (3) will not happen. A false positive will only occur under the circumstance that within the detection window, there exist $m$ consecutive packet collisions due to simultaneous transmissions. 

First, we present the mathematical model for the consecutive collision detector. Let $I_n$ be the indicator denoting the state of the $n$th observed packet transmission. Specifically,
\begin{equation*}
	I_n=\begin{cases}
		\;1,\qquad \text{the }n\text{th packet is a collision,} \\\; 0,\qquad \text{the }n\text{th packet is a successful transmission.}
	\end{cases}
\end{equation*}
We use $X_n$ to denote the state of the detector, then the behavior of the detector can be mathematically described as
\begin{align}
	\begin{cases}
		X_{n+1}=I_n\times (X_n+I_n)\\
		X_0=0.
	\end{cases}
	\label{detector}
\end{align}
With $m$ being the detection threshold, the decision rule of the detector in step $n$ is
\begin{equation*}
	\delta_n=\begin{cases}
		\;1 \quad \text{if} \quad X_n \geq m\\
		\;0 \quad \text{if} \quad X_n < m,
	\end{cases}
\end{equation*}

where $\delta_n$ is an indicator of whether $m$ consecutive collisions happen or not. The detector value $X_n$ will be reset to 0 as soon as it exceeds the threshold $m$ and the detection procedure starts over again.

Consider the sequence $\{X_n\}$ as a discrete random process, which takes values from a finite set $A=\{0,1,2,\dots,m\}$. The detector is said to be in state $i$ at step $n$ if $X_n=i$ with $i \in A.$ The state transition happens when a packet transmission over the wireless channel is observed. According to (\ref{detector}), the next state $X_{n+1}$ depends only on the current state $X_n$ and is independent of any other previous states, where the transition probability is 
\begin{equation*}
	P_{ij}=P\{X_{n+1}=j\;|\;X_n=i\}\quad i,j\in A.
\end{equation*}
Thus, the random process $\{X_n\}$ satisfies the Markov property and can be modeled as a discrete-time Markov chain. 

Given the decision threshold $m$, the Markov chain then can be described by the $(m+1)\times(m+1)$ transition probability matrix:
\[
\begin{bmatrix}
P_{00} & P_{01} &  \dots  & P_{0m} \\
P_{10} & P_{11} &  \dots  & P_{1m} \\
\vdots & \vdots &  \ddots & \vdots \\
P_{m0} & P_{m1} &  \dots  & P_{mm}
\end{bmatrix}
\]
Let $p_{ch}$ denote the channel collision probability, which is the probability that an observed packet transmission is a collision. Assuming the collision probability of each observed transmission is independent, then we have

\begin{equation}
	P_{ij}=\begin{cases}
		\;p_{ch}\qquad\qquad j=i+1\;\text{and }0\leq i\leq m-1\\
		\;1-p_{ch}\;\;\quad\quad j=0\;\text{and }0\leq i\leq m-1\\
		\;1\quad\qquad\qquad j=0 \;\text{and }i=m\\
		\;0 \quad\qquad\qquad \text{otherwise.}
	\end{cases}
\end{equation}

\begin{figure}[!ht]
	\centerline{\includegraphics[trim={0cm 23.5cm 5cm 2cm}, clip, width=3.7in]{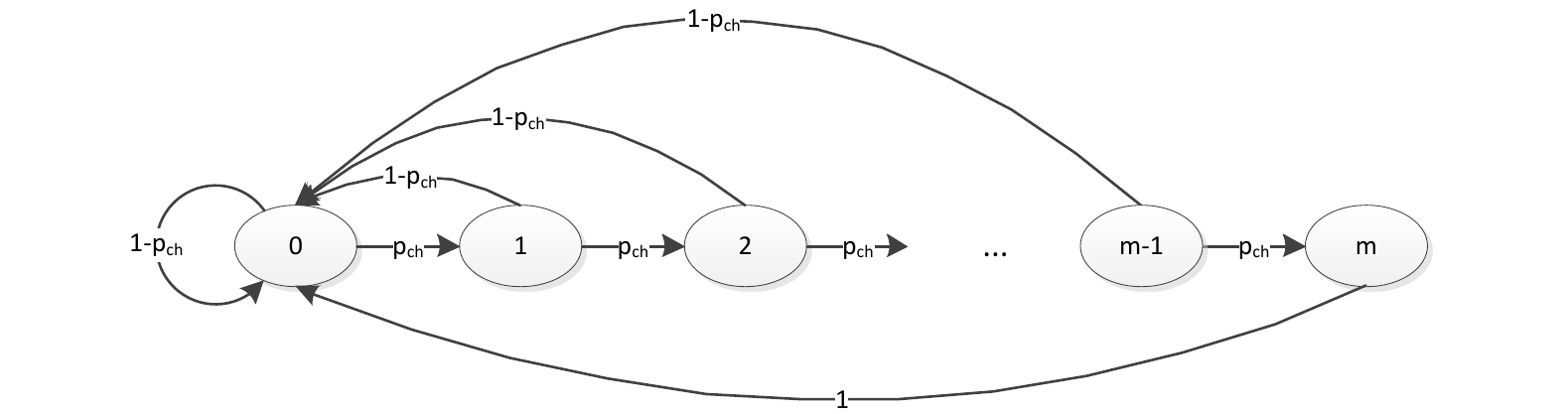}}
	\caption{Detector station transition diagram.}
	\label{markov}
\end{figure}

The state of the detector can only transit from state $i$ to state $i+1$ ($0\leq i\leq m-1$), or from state $i$ to state 0 ($0\leq i\leq m$).
The state transition diagram of the proposed detector is shown in Fig. \ref{markov}.

Let $(\pi_0,\pi_1,\dots,\pi_m)$ denote the steady-state probabilities of the Markov chain. $(\pi_0,\pi_1,\dots,\pi_m)$ can be solved from the following equations:

\begin{equation}
	\begin{cases}
		\;\pi_j=\sum\limits_{i=0}^{m}\pi_iP_{ij}, \qquad j\in\{0,1,\dots,m\}\\
		\;\sum\limits_{j=0}^{m}\pi_j=1.
	\end{cases}
\end{equation}

With equations (2) and (3), we can derive the close-form expression for $\pi_m$ as 
\begin{equation}
	\pi_m=\frac{p_{ch}^m-p_{ch}^{m+1}}{1-p_{ch}^{m+1}}.
\end{equation}

The detector will raise an alarm when it reaches state $m$. Under normal situations where no attacker is in presence, if the detector reaches state $m$ during the detection window, a false positive occurs. The false positive ratio (denoted as $P_{fp}$) of the proposed system is determined by the channel collision probability $p_{ch}$, the number of messages $m$ in the proposed key exchange protocol, as well as the number of packet transmission observed in the detection window. Assume there are total $k$ transmissions observed in the detection window, then $P_{fp}$ can be derived as
\begin{equation}
	P_{fp}=k\pi_{m}=k \cdot \frac{p_{ch}^m-p_{ch}^{m+1}}{1-p_{ch}^{m+1}}.
\end{equation}

From equation (5), we can see that given $k$ and $p_{ch}$, the false positive ratio of the proposed system monotonically decreases with $m$ increases. So we can always choose a large enough $m$ to reach an arbitrarily low target false positive ratio. It is worth noting that in implementation, $p_{ch}$ and $k$ will be estimated based on the transmission observed during the monitoring window $t$ (step 2 of Fig. \ref{fig:detector}) and may not be the exact packet collision probability for the transmissions observed during the detection window. Besides, equation (5) is derived under the assumption that the collision probability for each observed transmission is independent. So when implementing our system, the value of $m$ should be selected conservatively.

\subsection{Cost Analysis}
The cost introduced by applying the proposed solution has two aspects. First, the repeat transmission of the DH protocol message introduces extra communication overhead. Second, the shared key is considered to be valid at the end of the detection window, which results in a delay to the key establishment process.

The proposed solution repeats the DH protocol $m$ times, which has $2(m-1)$ more message transmission compared to the original DH protocol. Typically, the key agreement protocol is only used for initial trust establishment, and the subsequent key updates can be performed based on the existing shared secret. So the communication overhead of the proposed solution is in fact a one-time cost. From the numerical results in the next section, we can see that even at an extremely busy channel condition, the required value of $m$ to achieve a 1\% false positive ratio is no larger than 10. 

The delay introduced by the proposed solution equals to the key exchange timer $T$. $T$ consists of two windows: the channel monitoring window $t$ and the detection window $T-t$. The channel monitoring window is for Alice to estimate the channel collision probability and determine a value $m$ for the key agreement protocol, and the detection window should be large enough to cover the $2m$ protocol message transmission. The value of $T$ and $t$ is preset by Alice. In practice, we recommend using $T=1.5s$ and $t=1s$, since $1s$ monitoring window gives Alice plenty of samples for channel condition estimation, and a detection window of 0.5$s$ is enough for transmitting 26 maximum-sized protocol message with the proposed channel access mechanism under the lowest possible WiFi data rate (1Mbps). The recommended setting introduces a 1.5$s$ delay in the key establishment process, which is satisfactory considering that the required human interaction in existing OoB device pairing methods usually takes a couple of seconds.

\section{Numerical and Simulation Results}
In this section, we present the numerical and simulation results of the proposed system. We first present the simulation results of our collision detection algorithm, and compare the simulation results with theoretical analysis to demonstrate that our collision detection algorithm can precisely detect whether an observed transmission is a collision or a successful one. We then present the numerical and simulation results of the false positive ratio, and show that the proposed system can achieve an arbitrarily low false positive ratio. Last but not least, we demonstrate how to configure the system parameters through a case study.

In our simulation, Alice, Bob, and multiple background stations share an 802.11a wireless channel, with physical layer data rate configured to be OFDM 54Mbps. The packet size of the background traffic is uniformly distributed between 500 Bytes and 2000 Bytes. The link layer parameters are set to be the default values in the 802.11a standard. We summarize the key parameters of the simulation environment in Table \ref{table:simulation}.

\begin{table}
	\centering
	\begin{tabular}{ c||c }
		\hline
		Bit rate & 54Mbps \\
		Slot time & 9$\mu s$ \\
		DIFS & 34$\mu s$ \\
		SIFS & 18$\mu s$ \\
		ACK duration & 28$\mu s$ \\
		Initial backoff window size & 32\\
		Maximum backoff stages & 6 \\
		Maximum retry limit & 7 \\
		\hline
	\end{tabular}
	\caption{Simulation setup.}
	\label{table:simulation}
\end{table}

\subsection{Channel Collision Probability}
The channel collision probability under saturated traffic condition can be explicitly analyzed using the Markov chain based models\cite{bianchi2000performance,zhai2004performance}. We can use equation (14) and (17) in \cite{zhai2004performance} to calculate the channel access probability $\tau$ and the conditionally collision probability $p$. Then the channel collision probability $p_{ch}$ can be derived as

\begin{align}
	p_{ch}=&P\{\text{collision }|\text{ a transmission occurs}\} \nonumber \\
	=&\frac{P\{\text{collision}\}}{P\{\text{a transmission occurs}\}}\nonumber \\
	=&\frac{1-(1-\tau)^n-n\tau(1-\tau)^{n-1}}{1-(1-\tau)^n}
	\label{pch}
\end{align}

\begin{figure}[!ht]
	\centerline{\includegraphics[trim={0cm 0cm 0cm 0}, clip, width=2.3in]{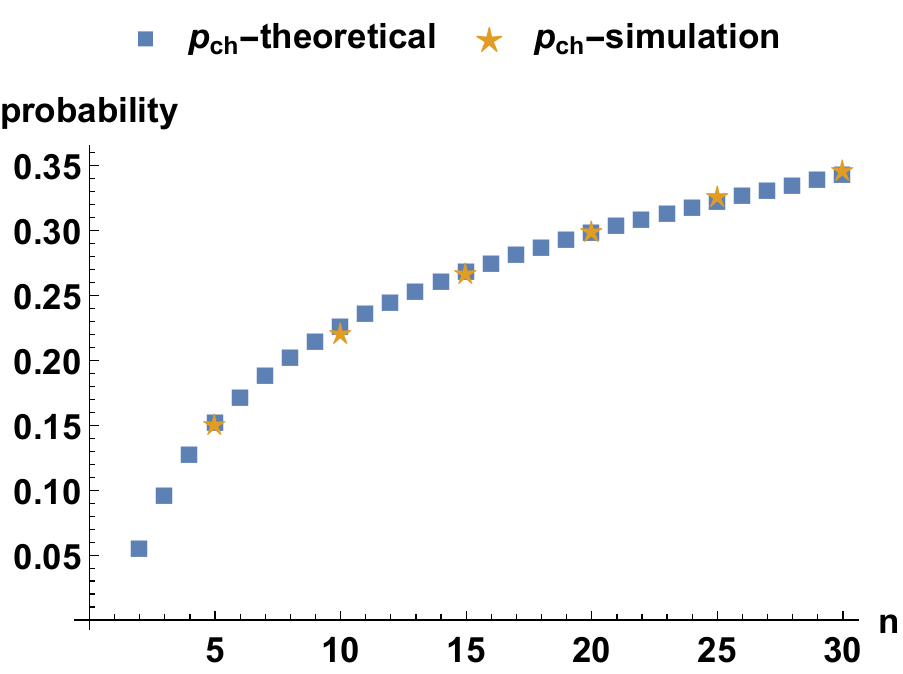}}
	\caption{Channel collision probability (saturated traffic).}
	\label{fig:pch}
\end{figure}

We numerically calculate $p_{ch}$ with the number of stations varies from 2 to 30, and plot the results in Fig. \ref{fig:pch}. We then implement the collision detection algorithm proposed in 4.4.2 on a silent node, which only monitors the channel without transmitting or receiving any packets. This silent node makes decisions purely based on the channel occupancy pattern it observed, and count the number of collisions and successful transmissions within the simulation run time. We plot the channel collision probability obtained from the silent node with the number of stations being 5, 10, 15, 20, 25, and 30 in Fig. \ref{fig:pch}. We can see that the $p_{ch}$ obtained by our collision detection algorithm matches the theoretical analysis with high precision.

\begin{figure}[!ht]
	\centerline{\includegraphics[trim={0cm 0cm 0cm 0}, clip, width=2.3in]{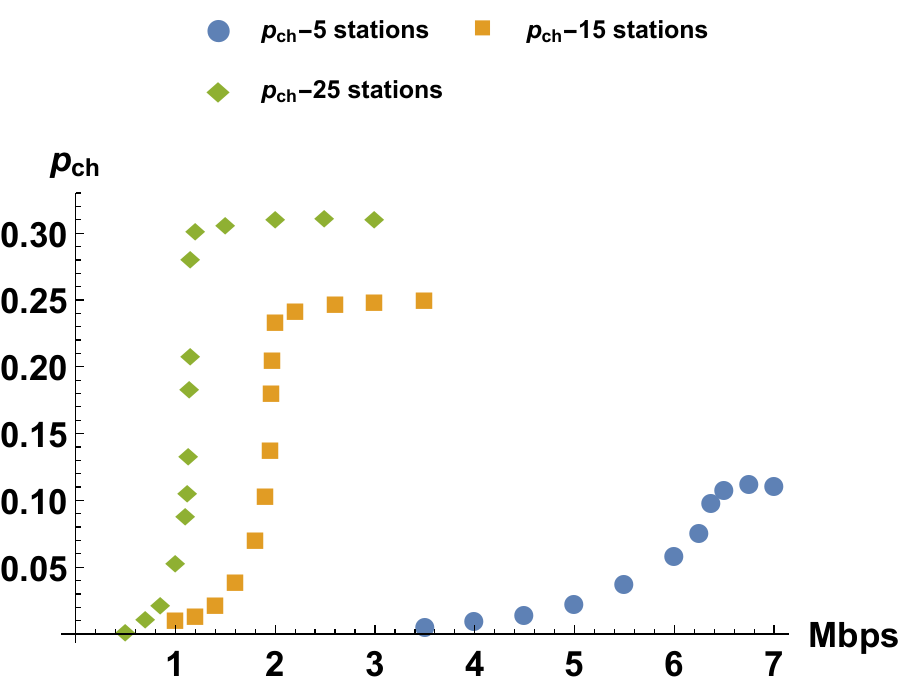}}
	\caption{Channel collision probability (unsaturated traffic).}
	\label{fig:uspch}
\end{figure}

We also implement the collision detection algorithm to obtain the channel collision probability $p_{ch}$ under unsaturated traffic conditions. To simulate the unsaturated traffic, we implement a Poisson traffic generator on each background station. Fig. \ref{fig:uspch} shows the $p_{ch}$ for 5, 15, and 25 stations under varies traffic densities. As long as the data rate of each station drops below a certain threshold, we can observe that $p_{ch}$ drastically decreases to a low level.

\subsection{False Positive Ratio}
\begin{figure}[!ht]
	\centerline{\includegraphics[trim={0cm 0cm 0cm 0}, clip, width=2.3in]{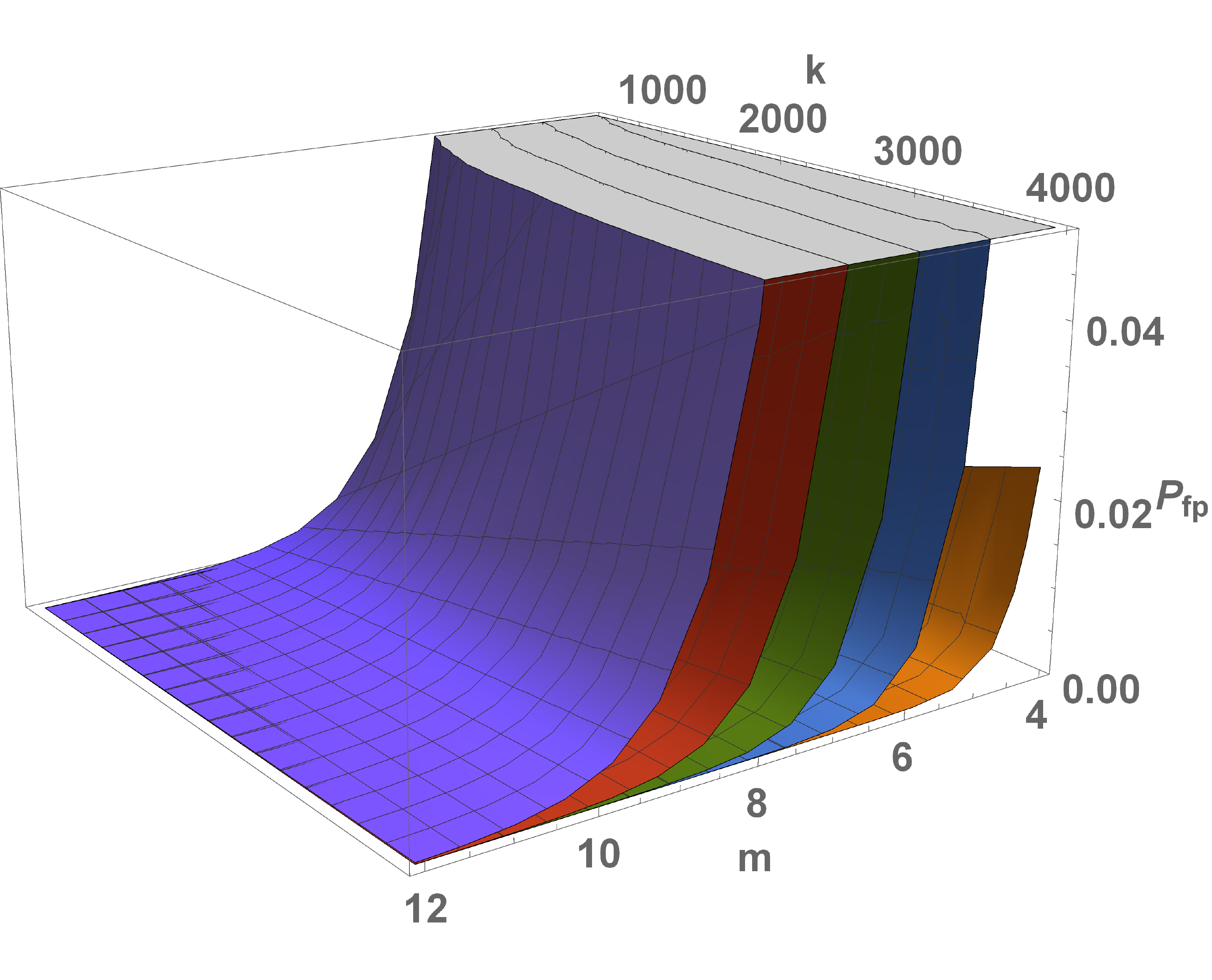}}
	\caption{False positive ratio.}
	\label{fig:pfp}
\end{figure}

As we analyzed in section 5.2, the false positive ratio of the proposed system is affected by a number of variables, including the channel collision probability $p_{ch}$, the number of transmissions being monitored within the detection window, as well as the number of messages transmitted in the proposed key exchange protocol $m$. By equation (5), $P_{fp}$ will increase with $p_{ch}$ and $k$ increasing, and with $m$ decreasing, which matches the intuition that these trends will give the normal packet transmission a better chance to hit $m$ collisions in a row. The numerical results of $P_{fp}$ is presented in Fig. \ref{fig:pfp}. The five surfaces from bottom to top represent the $P_{fp}$ with $p_{ch}$ being 5\%, 10,\% 15\%, 20\%, and 25\%, respectively, while $k$ varying from 1000 to 4000 and $m$ ranging from 4 to 12. When $p_{ch}$ equals 25\%, which means the channel is operating in an extremely busy condition, the proposed system can achieve a false positive ratio of 1\% if $m$ is larger than 10.

We conduct extensive simulations in both saturated and unsaturated traffic scenarios to examine the false positive ratio of the proposed system. First, we set 5 stations to generate saturated traffic, and have one silent node to monitor the channel and record the consecutive collisions it observed. The detection window is set to be 0.5 seconds, and on average there are 1545 transmissions observed in one detection window. We conduct the simulation 20000 times, and the results are presented in Table \ref{table:pfpsaturated}. In this situation, the proposed scheme can achieve 1\% false positive ratio with $m=5$. The $P_{fp}$ obtained in the simulation is lower than the theoretical result in equation (5). This is because equation (5) is based on the assumption that each packet collision is independent. However, after a collision happens, the involved stations will double their contention window for the next transmission, so the collision probability for the next transmission will be smaller. So the $P_{fp}$ calculated by equation (5) is in fact an upper bond for the saturated traffic scenario.

\begin{table}
	\centering
	\begin{tabular}{ |c|c| c| c|}
		\multicolumn{4}{c}{5 stations $\;\;$   detection window: 0.5$s$ $\;\;$    simulation run: 20000}\\
		\hline  \hline
		$m$ & No. alarms & $P_{fp}$ & 95\% confidence interval\\
		\hline
		4 & 446 & 2.23\% & $(1.76\%,\; 2.70\%)$\\
		\hline
		5 &  17 &  0.085\%& $(0.007\%,\;0.177\%)$\\
		\hline
	\end{tabular}
	\caption{False positive ratio (saturated).}
	\label{table:pfpsaturated}
\end{table}

The simulation results for unsaturated traffic is presented in Table. \ref{table:pfpunsaturated}. 12 stations are equipped with Poisson traffic generators with data rate being 1.875Mbps. On average, there are 1198 transmissions observed during the 0.5$s$ detection window. The simulation results show that the actual false positive ratio is higher than the theoretical value calculated using equation (5). This is due to the fact that under the unsaturated condition, most of the times only a few stations simultaneously have packets ready to transmit, resulting in a low average channel collision probability. However, when a collision happens, the involved stations will attempt to retransmit their packets, as a result, the number of stations competing for the next transmission is higher than average, which leads to a higher than average channel collision probability. So when implementing our system in an unsaturated condition, Alice should select $m$ conservatively. We recommend to first calculate $m$ with Equation (5) to obtain the value required for a target false positive ratio, then use $m+2$ in the proposed solution for a guaranteed performance.
\begin{table}
	\centering
	\begin{tabular}{ |c|c| c| c|}
		\multicolumn{4}{c}{12 stations $\;\;$1.875 Mbps$\;\;$    $\;\;$    simulation run: 20000}\\
		\hline  \hline
		$m$ & No. alarms & $P_{fp}$ & 95\% confidence interval\\
		\hline
		4 & 932 & 4.66\%   & $(3.99\%,\; 5.33\%)$\\
		\hline
		5 &  126 &  0.63\%& $(0.38\%,\;0.88\%)$\\
		\hline
		6 &   15   &    0.075\%    &$( 0,\;0.16\% )$    \\
		\hline
	\end{tabular}
	\caption{False positive ratio (unsaturated).}
	\label{table:pfpunsaturated}
	\vspace{-0.2cm}
\end{table}

\subsection{A Case Study}

In this subsection, we present a simulation case to demonstrate how does Alice select the system parameter $m$ based on the information she observed during the monitoring window, to achieve a target false positive ratio. In this simulation case, Alice, Bob, and other 10 background stations are sharing an 802.11a wireless channel. Each of the 10 background stations is generating packets with 2.0 Mbps data rate. The key exchange timeout $T$ is set to be 1.5 seconds, and the duration of the monitoring window $t$ is set to be 1 second. Based on the information observed during $t$, Alice will decide $m$, which is the number of messages to be transmitted in the proposed key exchange protocol, to reach a target $P_{fp}$ of 0.5\%.

After receiving the association reply from Bob, Alice starts to monitor the channel for $t=1$ second. During this time, Alice observed 2065 transmission events, among them there are 1994 successful transmissions and 71 collisions (these numbers are obtained from a simulation case). Based on these values, Alice estimates the channel collision probability $p_{ch}$ to be 3.44\%, and the number of transmissions in the detection window ($T-t=0.5$ seconds) to be 1033. According to equation (5), if $m$ equals 4, $P_{fp}$ is 1.36\%, and if $m$ takes the value 5, $P_{fp}$ is estimated to be 0.08\%, which reaches the target false positive ratio. As we mentioned earlier, $m$ should be selected conservatively. So Alice will set $m$ to be 7 and starts the key exchange protocol. 

We conduct simulations for both the normal case and the MITM attack scenario. We let Alice transmit 7 maximum-sized packets at the beginning of the detection window with her backoff counter fixed to 0. The attacker is set to transmit a jamming packet without channel sensing and backoff process, and the transmission of the attacker is triggered when Alice starts to transmit. In the normal case, during the 0.5 seconds detection window, Bob observes 41 collisions and 996 successful transmissions on the channel, and the maximum length of consecutive collisions observed is 2. In the attack case, Bob successfully detects 7 consecutive collisions at the very beginning of the detection window, which indicates the presence of the attacker.

\section{Dsicussion: MITM Attack vs Impersonation Attack}

The proposed scheme is based on the assumption that Alice and Bob have already gone through an association handshake and have a wireless link available between them. However, the attacker may impersonate Alice or Bob during the association phase. Specifically, when Alice sends the association request, the attacker can jam this message and then impersonate Bob to send a reply to Alice. The proposed scheme covers the key exchange process after the association, but not the association phase itself. The essential difference between the impersonation attack and the MITM attack is that in the latter case, both Alice and Bob know that a key exchange process is expected within a certain time window (corresponding to the key exchange timeout window $T$ in the proposed solution). However, in the former case, being the passive party in the association handshake, Bob is not aware of that Alice is trying to establish a shared key with him.

The proposed solution can work on top of any existing identity authentication technique, such as the push button configuration (PBC) mechanism defined in 802.11 standards, to address the impersonation attack. However, an identity authentication protocol requires either pre-shared knowledge or OoB channels. In fact, we can address the impersonation attack by a similar link layer approach. Due to the space limitation, here we only introduce the general methodology for a secure association protocol without going deep into the detailed design.

In the secure association protocol, Alice transmits the association request $m$ times, with random delays between each request, and Bob replies with $n$ association replies once he receives an association request. The $n$ association replies will be transmitted using the channel access mechanism proposed in section 4.3. To successfully impersonate Bob, the attacker has to jam all the $m$ requests from Alice, otherwise if Bob receives a single request, the attacker has to jam the corresponding $n$ replies, which results in $n$ consecutive collisions at Alice's side and can be detected by Alice.  With the random delay between the association requests, even if the attacker knows the exact transmission starting time of the first request, he does not know the transmission starting time of the following requests. The best he can do is to examine the packet header of every transmission on the channel, and sending a jamming signal when he sees the current packet is indeed from Alice. If the packet header containing Alice's address is perceptible to the attacker, it is also perceptible to Bob. Based on this insight, Bob can keep monitoring the channel and keep a record on the source address of collided transmissions. If the attacker jams all the $m$ requests from Alice, Bob will notice that $m$ transmissions from the same address collided consecutively. At this point, Bob sends $n$ alarm messages using the channel access scheme in section 4.3. If Alice receives an alarm message, she will be aware of the attacker; if all the alarms are jammed by the attacker, Alice will observe $n$ consecutive collisions and be aware of the attacker.

\section{Conclusion}
In this paper, we systematically studied the wireless MITM attack, and modeled the attacker's behavior on message level. We then present a novel in-band solution to detect the attacker during the key exchange process. The proposed scheme forces the attacker to generate a burst sequence of consecutive channel collisions, which can be detected by legitimate parties. We further present the theoretical analysis as well as the numerical and simulation results to validate the performance of the proposed system. Our solution achieves a missed detection ratio of 0, and can achieve an arbitrary target false positive ratio by proper parameter design. A case study is also presented to demonstrate how to configure the system to achieve a guaranteed performance.

\bibliographystyle{IEEEtranS}

\end{document}